\documentclass[pdflatex,sn-mathphys-num]{sn-jnl}

\newcommand{\araa}{Annu. Rev. Astron. Astrophys.}   
\newcommand{\apj}{Astrophys. J.}   
\newcommand{\apjl}{Astrophys. J. Lett.}   
\newcommand{\apjs}{Astrophys. J. Suppl. Ser.}   
\newcommand{\aap}{Astron. Astrophys.}   
\newcommand{\aapr}{Astron. Astrophys. Rev.}   
\newcommand{\aaps}{Astron. Astrophys. Suppl.}   
\newcommand{\mnras}{Mon. Not. R. Astron. Soc.}   
\newcommand{\nat}{Nature} 
\newcommand{\pasa}{Publ. Astron. Soc. Aust.}   
\newcommand{\pasp}{Publ. Astron. Soc. Pac.}   

\newcommand{\kms}{km\,s$^{-1}$}

\newcommand{\lya}{Ly$\alpha$}

\newcommand{\foiii}{\hbox{[O \textsc{III}]}}

\newcommand{\hb}{\hbox{H$\beta$}}

\newcommand{\farcs}{$.\!\!^{\prime\prime}$}

\usepackage{graphicx}%
\usepackage{multirow}%
\usepackage{amsmath,amssymb,amsfonts}%
\usepackage{amsthm}%
\usepackage{mathrsfs}%
\usepackage[title]{appendix}%
\usepackage{xcolor}%
\usepackage{textcomp}%
\usepackage{manyfoot}%
\usepackage{booktabs}%
\usepackage{algorithm}%
\usepackage{algorithmicx}%
\usepackage{algpseudocode}%
\usepackage{listings}%
\usepackage[normalem]{ulem}
\graphicspath{{./figs/}}
\usepackage{rotating}

\newcommand{\editone}[1]{#1}

\begin{document}

\title[Article Title]{Extended Enriched Gas in a Multi-Galaxy Merger at Redshift 6.7}

\author*[1,2]{\fnm{Weida} \sur{Hu}} \email{weidahu@tamu.edu}
\author[1,2]{\fnm{Casey} \sur{Papovich}}
\author[1,2]{\fnm{Lu} \sur{Shen}}
\author[3]{\fnm{Zixuan} \sur{Peng}}
\author[4]{\fnm{L. Y. Aaron} \sur{Yung}}
\author[5]{\fnm{Brian C.} \sur{Lemaux}}
\author[1,2]{\fnm{Justin} \sur{Spilker}}
\author[1,2]{\fnm{Justin} \sur{Cole}}

\affil*[1]{\small \orgdiv{Department of Physics and Astronomy}, \orgname{Texas A\&M University}, \orgaddress{\city{College Station}, \state{TX} \postcode{77843-4242}, \country{USA}}}
\affil[2]{\small George P. and Cynthia Woods Mitchell Institute for Fundamental Physics and Astronomy, Texas A\&M University, College Station, TX 77843-4242, USA}
\affil[3]{\small Department of Physics, University of California, Santa Barbara, Santa Barbara, CA 93106, USA}
\affil[4]{\small Space Telescope Science Institute, 3700 San Martin Drive, Baltimore, MD 21218, USA}
\affil[5]{\small Gemini Observatory, 670 N. A’ohoku Place, Hilo, Hawai'i, 96720, USA}

\abstract{\bf Recent JWST observations have uncovered high-redshift galaxies characterized by multiple star-forming clumps, many of which appear to be undergoing mergers \cite{Hashimoto2023,Hainline2024,Chen2023,Arribas2024}.
Such mergers, especially those of two galaxies with equivalent masses, play a critical role in driving galaxy evolution \cite{DiMatteo2008,Moreno2019,Xie2024,Turner2025,Duan2024} and regulating the chemical composition of their environments \cite{Hani2018,Sparre2022}. 
Here, we report a major merger of at least five galaxies, dubbed JWST's Quintet (JQ), at redshift 6.7, discovered in the JWST GOODS-South field.
This system resides in a small area $\sim4.5''\times4.5''$ ($24.6\times24.6$ pkpc$^2$), containing over 17 galaxy-size clumps with a total stellar mass of $10^{10}\ M_\odot$.
The JQ system has a total star formation rate of 240 -- 270 $M_\odot$ yr$^{-1}$, placing it $\sim1$ dex above the median star formation rate--mass main sequence at this epoch.  
The high mass and star formation rate of the JQ galaxies are consistent with the star formation history of those unexpected massive quiescent galaxies observed at redshift 4 -- 5 \cite{Carnall2023,Nanayakkara2024,deGraaff2024}, offering a plausible evolutionary pathway for the formation of such galaxies.
We also detect a large \foiii+H$\beta$ emitting gaseous halo surrounding and connecting four galaxies in the JQ, suggesting the existence of heavy elements in the surrounding medium--inner part of its circumgalactic medium (CGM). 
This provides direct evidence for the metal enrichment of galaxies' environments through merger-induced tidal stripping, just 800 Myr after the Big Bang.}

\maketitle

\begin{figure*}
    \centering
    \includegraphics[width=1\linewidth]{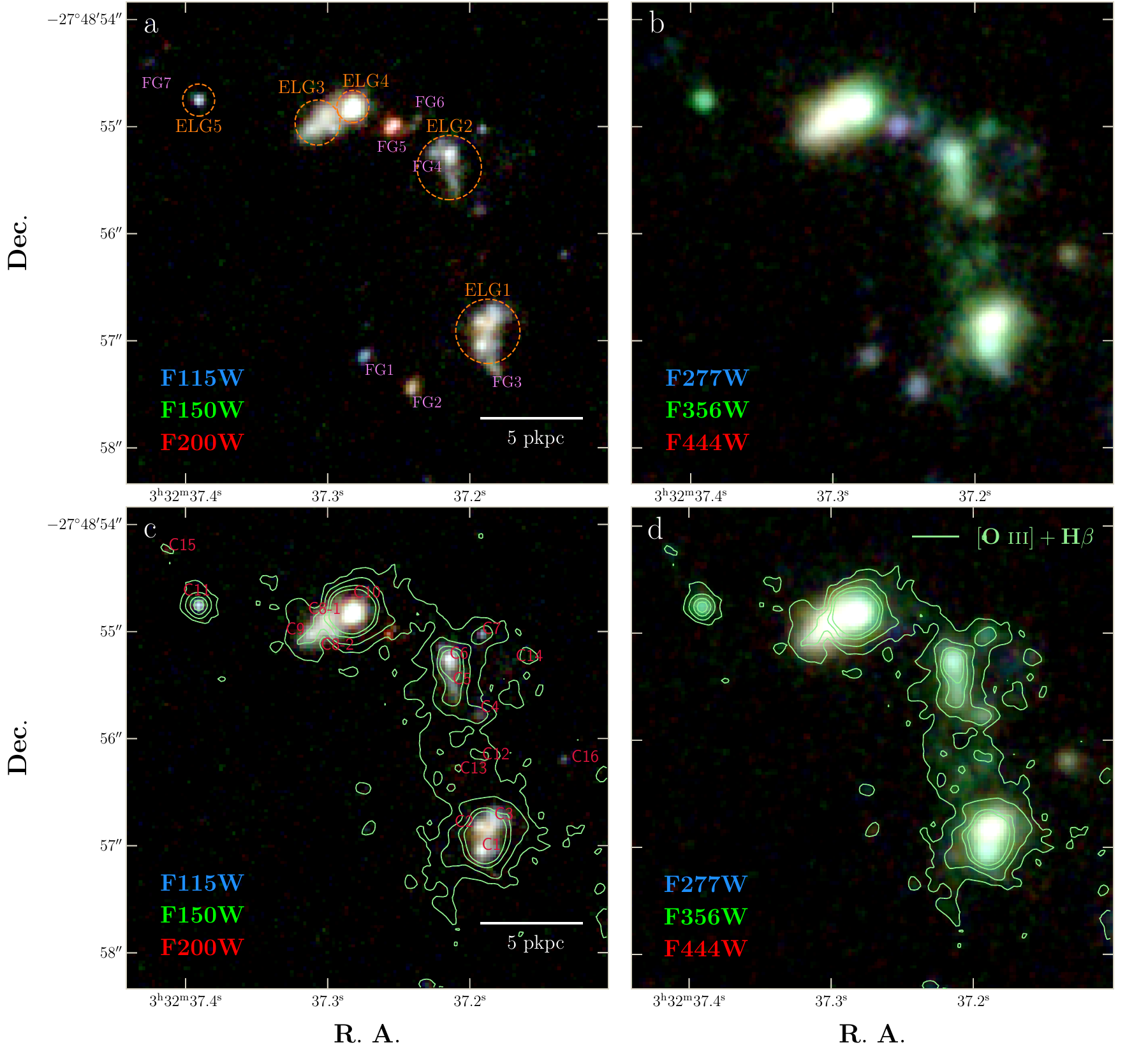}
    \caption{\textbf{The pseudo-color images of JWST's Quintet at redshift 6.71.} \textbf{a)~} The pseudo-color image created from NIRCam F115W (blue), F150W (green), and F200W (red) images. 
    The five emission line galaxies in JWST’s Quintet (JQ) are indicated by large orange circles and labeled (ELG1--ELG5). Small purple labels denote seven foreground galaxies at different redshifts and are unassociated with the JQ. The white bar in the lower right shows a physical scale of 5 pkpc.
    \textbf{b)~} The pseudo-color image created from NIRCam F277W (blue), F356W (green), and F444W (red) images. The F356W (green) filter contains the diffuse emission of redshifted \foiii+\hb, which is visible in the pseudo-color image.
    \textbf{c)~} Similar to \textbf{a}, but with the foreground galaxies removed. Red labels (C1 -- C16) mark clumps belonging to the JQ galaxies. 
    \textbf{d)~} Similar to \textbf{b}, but with the foreground galaxies removed. 
    In panels \textbf{c} and \textbf{d} the diffuse \foiii+\hb\ halo is shown by the green contours. 
    We estimate the stellar continuum of clumps from the SED fitting and subtract their contribution from the F356W image to obtain the \foiii+\hb\ contour map of JQ (see Methods).
    Here, we present the smoothed \foiii+\hb\ map using a Gaussian kernel with a width of 1 pixel (0.03 arcsec). In Extended Data Figure \ref{fig:oiiimap}, we present the unsmoothed \foiii+\hb\ map.
    \editone{We show contour levels 2, 5, 10, and 15 $\sigma$, which indicate surface brightnesses of [8.8, 22, 44, 66] $\times10^{-17}$ erg s$^{-1}$ cm$^{-2}$ arcsec$^{-2}$.}
    }
    \label{fig:rgb}
\end{figure*}

\begin{figure*}
    \centering
    \includegraphics[width=\linewidth]{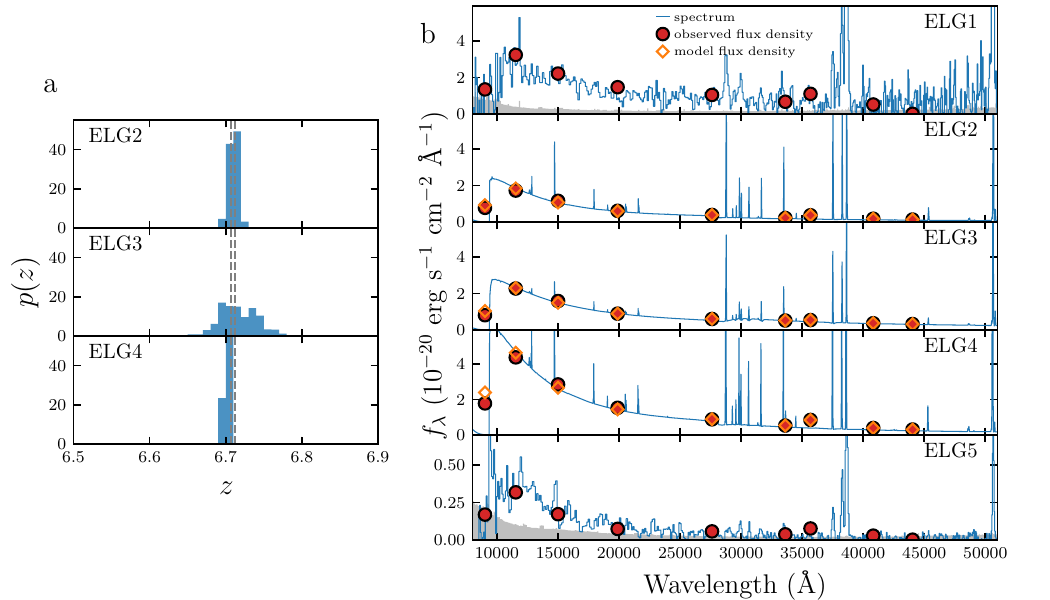}
    \caption{\textbf{Redshift measurements and SEDs of the ELGs in JQ.} \textbf{a)~} Photometric-redshift probability-distribution functions, $p(z)$, for ELG2, ELG3, and ELG4 \editone{(zoom into the redshift range of 6.5--6.9)}. Although these galaxies lack spectroscopic data, their $p(z)$ are centered at the spectroscopic redshift of ELG1 and ELG5, indicated by the vertical dashed lines, with very small uncertainties.
    \textbf{b)} The SEDs of all JQ ELGs.  The red dots show the flux densities from the broad- and medium-band imaging filters. 
    For ELG1 and ELG5, the blue curves show the NIRCam Prism spectra from JADES, with the correction factors applied to account for JWST/NIRSpec slitlosses (see Methods). 
    For ELG2, ELG3, and ELG4, the blue curves show the best-fit model spectrum and the orange diamonds show the corresponding model flux densities in the imaging bands.  The SEDs are highly similar, consistent with their being at the same redshift.}
    \label{fig:photoz}
\end{figure*}

We identified a group of five emission line galaxies (ELGs) at redshift 6.7, which we call JWST's Quintet (JQ), from a study of ELGs based on imaging and spectroscopic data from the JWST Advanced Deep Extragalactic Survey \cite{Rieke2023} (JADES).
JADES, one of the deepest JWST/NIRCam imaging surveys to date, provides continuous wavelength coverage from 0.8 to 5.0 $\mu$m across 14 broad- and medium-band filters (see Methods).
\editone{The JADES multi-wavelength imaging data enables precise photometric redshift estimation with an uncertainty of $\Delta z/(1+z) = 0.024$} \cite{Rieke2023}, allowing reliable distances to galaxies and distinguishing galaxies in close physical proximity.
We identified ELGs with strong \foiii+\hb\ nebular emission as well-detected ($>10\sigma$) sources with photometric redshifts in the range 6.7 to 7.6 with a $2\sigma$ excess of flux emission in the F410M medium-band image compared to the F444W broad-band image of JADES.   At these redshifts, the \foiii\ and \hb\ emission lines fall in the F410M bandpass, and the flux excess criteria identify objects with EW(\foiii+\hb) $>$ 300~\AA.  
The spatial distribution of ELGs revealed an overdensity of the five galaxies within a $4.5''\times4.5''$ region, see Figure \ref{fig:rgb}, corresponding to $\sim 25\times25$ proper kpc$^{2}$ (pkpc$^2$)
Of these, two galaxies (ELG1 and ELG5) have spectroscopically confirmed redshifts of 6.707 and 6.712 based on the NIRSpec spectroscopy \cite{D'Eugenio2024}.
These two galaxies reside at the head and the tail of the JQ with a projected distance of 18.6 pkpc and a velocity difference of 195 \kms. The remaining three ELGs (ELG2, ELG3, ELG4) reside between them and have similar spectral energy distributions (SEDs) as ELG1 and ELG5 with consistent photometric redshifts $6.70\pm0.03$ (see Figure \ref{fig:photoz}). This is evidence that all five ELGs are in the same physical structure.

The JQ appears to be a high-redshift \editone{situation similar to that in the well-known galaxy group}, Stephan's Quintet, at a distance of 85 Mpc. 
Unlike Stephan's Quintet, where only four of the five galaxies are physically associated and interacting in mergers \cite{Moles1997,Appleton2023}, we believe all galaxies in JQ are physically associated.
Furthermore, there is evidence that the entire structure of ELG1, ELG2, ELG3, and ELG4 is connected by a diffuse emission of \foiii+\hb\ only seen in the F356W image. 
At the redshift of JQ, the \foiii~$\lambda5007$ line falls within the F410M band, facilitating the detection of this system, but the F356W band encompasses both \foiii\ and \hb\ with much higher transmission than F410M.  We therefore use F356W for analysis in this paper.
The implication is that these galaxies have undergone an interaction that produced the diffuse \foiii+\hb\ emission.
In ELG2, a notably extended tail (labeled C5 in Figure~\ref{fig:rgb}) points toward ELG1 in the projected plane, with several faint star-forming clumps (C4, C12, C13; clump identification is described below) distributed along it. These clumps likely trace the star formation induced by the galaxy mergers.
ELG3 also shows a disturbed morphology with two clumps (C8-1, C8-2) connected through diffuse emission to ELG4. 
\editone{The velocity offset (195 \kms) between ELG5 and ELG1, and the projected distance (6.2 pkpc) between ELG5 and ELG3, are both smaller than the virial velocities and radii of these ELGs (see Methods), suggesting that ELG5 is gravitationally bound to the JQ system.   
Furthermore, ref \cite{Witstok2024} reported an unusually high \lya\ escape fraction ($55\pm4\%$) in ELG5, suggesting it resides within a large ionized bubble -- too large to be produced by ELG5 alone. 
Generating such an ionized bubble requires substantial contributions from the entire JQ system. 
These facts support that ELG5 is associated with the JQ.
However, we caution that the degeneracy between line-of-sight distance and velocity offset introduces uncertainty; it remains possible that the true three-dimensional separation between ELG5 and the other ELGs exceeds the virial radius, in which case ELG5 would not yet be bound.}

The JQ galaxies show a high frequency of spatially resolved clumps.  
We adopt an automatic clump identification algorithm combined with visual inspection to detect the clumps in this structure.  We measure the clump sizes and photometry by fitting their surface brightness with S\'ersic profile models, and we measure the galaxy properties by fitting models to the SEDs. 
We identified 17 clumps associated with the JQ and 7 foreground galaxies.
Extended Data Table \ref{tab:properties} lists the properties of the JQ clumps.
To account for the diffuse emission surrounding the clumps, we also perform a spatially resolved SED fitting to measure the integrated properties of the entire system. 
We derive a total stellar mass of $9.9^{+0.7}_{-0.4}\times10^9$ $M_\odot$ and star formation rate (SFR) of $255^{+13}_{-14}$ $M_\odot$ yr$^{-1}$.
We remove the foreground galaxies from the images based on their best-fit S\'ersic profiles and present a clean map of JQ in Figure \ref{fig:rgb}c,d. See Methods for details.

The sizes and SFRs of the JQ clumps are significantly larger than those of the star clusters and star cluster-size clumps within individual galaxies resolved through the lensing technique \cite{Adamo2024,Vanzella2023,Fujimoto2024}.
Their sizes, stellar masses, and SFRs are more consistent with the high-redshift spatially unresolved field galaxies observed with JWST NIRCam \cite{Morishita2024}. 
\editone{We, therefore, define them as galaxy-size clumps.}
Although recent JWST observations have reported numerous $z>6$ galaxy systems containing such galaxy-size clumps \cite{Chen2023,Hainline2024,Jones2024,Hashimoto2023}, JQ stands out as the first system identified with over 10 galaxy-size clumps.
This high density of galaxy-size clumps within a single galaxy system is extremely rare and challenges predictions from cosmological simulations.
For example, work focusing on clump formation during mergers between high-redshift galaxies does not identify any systems with more than four clumps at $z > 6.5$ \cite{Nakazato2024}.
Comparing JQ to predictions from the $z\sim$ 6.5 -- 7.0 galaxy systems in a 2-deg$^2$ simulation \cite{Yung2023}, we identify only 8 systems with $\geq10$ galaxy-size clumps out of a parent sample of 29,696 systems \editone{, and 4 of which are classified as major mergers}.   From this simulation, we would expect only $\simeq$ \editone{0.13} JQ-like systems in the JADES survey at $z\sim 6.7$ -- 7.6.
This indicates the number density of galaxy-size clumps in the JQ is higher than $>99.97\%$ galaxy systems in cosmological simulation, and is a rare configuration.  
The small area of JADES F410M image ($\sim120$ arcmin$^2$) implies an observed comoving number density of $3.7^{+3.1}_{-3.0} \times 10^{-6}$ cMpc$^{3}$ for JQ, almost \editone{one dex larger than} the expected number density from the cosmological simulation ($4.7 \times 10^{-7}$ cMpc$^3$). See Methods for details.

\begin{figure}
    \centering
    \includegraphics[width=0.8\linewidth]{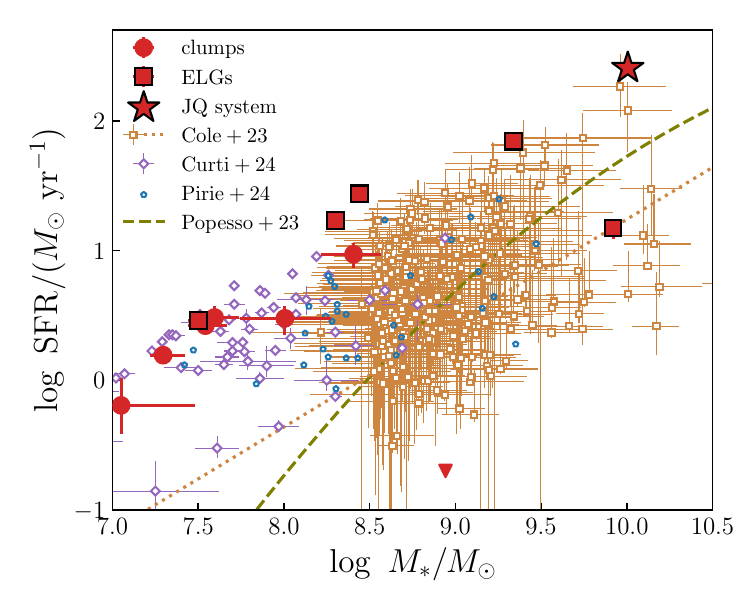}
    \caption{\textbf{Location of the five galaxies (red squares), the associated clumps (red dots), and the entire JQ system (red star) in the star formation rate -- stellar mass plane.} The clump C16 with a very small SFR of $0.005^{+0.057}_{-0.005}\ M_\odot$ yr$^{-1}$ is marked by a red triangle.  
    We also plot a series of measurements at similar redshifts from the literature and the best-fit SFMS relations based on those samples \cite{Cole2023,Popesso2023,Curti2024,Pirie2024}.
    In all cases, the clumps in JQ with stellar masses $\log M_\star / M_\odot \lesssim 9$ have SFRs more than 1 dex above the measured SFMS relation of ref \cite{Cole2023}, indicating the clumps are in a bursting phase.
    The total SFR of galaxies in the JQ system is higher than the SFMS by $\sim 1$ dex and among the top SFR of the galaxies at similar redshifts. \editone{In Extended Data Figure \ref{fig:mass_sfr}, we present the same figure, but with the five galaxies replaced by their member clumps.}}
    \label{fig:mass_sfr_galaxy}
\end{figure}

\begin{figure}
    \centering
    \includegraphics[width=0.9\linewidth]{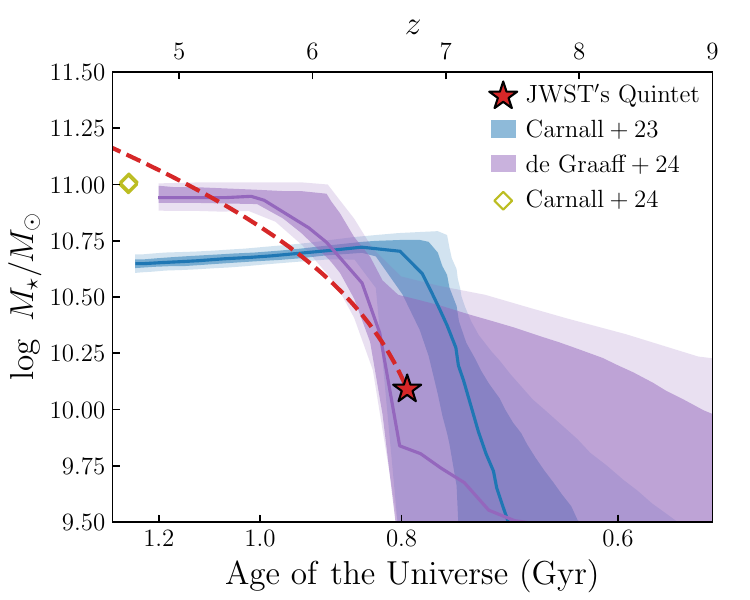}
    \caption{\textbf{The expected stellar mass growth of the JQ system (red dashed line).} We overlay the star formation histories of two massive quiescent galaxies at $z\sim4.5$ -- 5, recently discovered with JWST \cite{Carnall2023,deGraaff2024}, that have SFHs similar to the extrapolated growth of JQ. The shaded regions indicate the 1$\sigma$ and 2$\sigma$ confidence intervals of their star formation histories. We also include two newly confirmed massive quiescent galaxies from ref \cite{Carnall2024} as the olive-green diamonds. The future growth of the JQ system is consistent with these tracks.}
    \label{fig:massgrowth}
\end{figure}

\editone{The five galaxies and their associated clumps show evidence of elevated SFRs compared to typical galaxies at these epochs. 
Figure \ref{fig:mass_sfr_galaxy} compares the SFRs of the clumps to the galaxies at similar redshift and the SFR--Mass Main Sequence (SFMS) using results from the literature \cite{Cole2023,Curti2024,Pirie2024,Popesso2023}. 
All five galaxies and clumps, except ELG3 and C16, have SFRs more than 1 dex above the observed SFMS relation \cite{Cole2023}, and the total SFR of JQ is also higher than the SFMS by $\sim1$ dex, indicating they are in a bursting phase.
Best-fit star-formation histories show that these galaxies and clumps could form all of their stellar mass in the past 10 Myr, but this is a lower limit, as the recently formed stars could outshine previous generations of stars \cite{Conroy2013,Papovich2023}. 
Nevertheless, the short formation timescales imply that the star formation in these galaxies and clumps was recently triggered by major mergers in JQ \cite{Nakazato2024}.
In contrast to those less massive components, the most massive galaxy, ELG3 with a stellar mass of $9.92\ M_\odot$, contributing $>80\%$ of the total stellar mass, shows no evidence of an enhanced SFR. Its SFR accounts for $<10\%$ of total SFR and lies on the median SFMS relation at this epoch. 
This suggests that this merger has impacted the star formation of the lower-mass galaxies and clumps more than ELG3. 
Consequently, the future mass growth of JQ will be driven by the combination of its star formation along the main sequence and its merging companions, consistent with the ex-situ star formation scenario suggested for massive major mergers \cite{Tacchella2019}.}

The \editone{substantial \textit{ex-situ} star formation and} elevated SFR triggered by major mergers in JQ provides a plausible evolutionary pathway for forming massive quiescent galaxies, recently discovered at $z\sim$ 4.5 -- 5. The existence of these objects has challenged our understanding of rapid galaxy growth and quenching over short timescales \cite{Carnall2023,deGraaff2024,Carnall2024,Glazebrook2024}. 
The high stellar mass and SFR of the JQ system are consistent with those needed to produce massive ($>10^{10.5}\ M_\odot$), quiescent galaxies observed at redshift 4 -- 5 \cite{Carnall2023,deGraaff2024} (see Figure \ref{fig:massgrowth}).
The expected merging timescale for the JQ system is $\Delta t \sim100$ -- 350 Myr (see Methods).
If the JQ galaxies sustain the current SFR during this period, they will build a single massive galaxy with a stellar mass of SFR $\times \Delta t \sim 10^{10.5}$ -- $10^{11}\ M_\odot$.
Simulations show that such a major merger event is expected to rapidly consume cold gas and trigger strong feedback from black hole activities, ultimately quenching star formation in the JQ galaxies \cite{Mihos1996,DiMatteo2005,Hopkins2008}.
Furthermore, the observed number density of JQ-like systems is consistent with that of massive quiescent galaxies at redshift $\sim5$, estimated to be $3-4.5\times10^{-6}$ Mpc$^{-3}$ \cite{deGraaff2024,Carnall2023}. 
These factors suggest that JQ is a likely candidate for a progenitor of the massive quiescent galaxies identified at redshifts 4 -- 5.  To test this interpretation requires a detailed study of the kinematics of the individual JQ clumps.
\editone{This can be achived with future observations from JWST NIRSpec integral field unit or NIRCam wide field slitless spectroscopy (WFSS).
In particular, a NIRCam WFSS observation using the F356W filter is expected in fall 2025 \cite{Rieke2023b},  which will provide the kinematics and \foiii+\hb\ fluxes of bright clumps in JQ.}

The NIRCam F356W image of JQ is dominated by the redshifted \foiii+\hb\ emission line.  The image shows extended, diffuse emission, compared to the NIRCam F335M and F277W images, which is evidence of a \foiii+\hb\ halo surrounding the JQ system.
We estimate the amount of \foiii+\hb\ emission by subtracting the continuum contribution from the F356W image (see Methods) and present the \foiii+\hb\ halo as the green contours in Figure \ref{fig:rgb} (the original \foiii+\hb\ map is shown in Extended Data Figure \ref{fig:oiiimap}).
The \foiii+\hb\ halo surrounds the entirety of the ELG1, ELG2, ELG3, and ELG4 galaxies, extending more than 20~kpc.  The total \foiii+\hb\ luminosity is $10^{43.77\pm0.10}$ erg s$^{-1}$.
Notably, we observe a ``bridge'' in the \foiii+\hb\ emission connecting galaxies ELG1 and ELG2.
This bridge extends to 5 kpc and includes the clumps C12 and C13. 
The presence of this large-scale \foiii+\hb\ halo requires oxygen-enriched gases in the \editone{surrounding medium of galaxies, the inner part of the CGM.
These oxygen-enriched gases were} ejected and/or stripped efficiently from the galaxies in JQ, particularly from ELG1 and ELG2, as evidenced by the bright \foiii+\hb\ bridge connecting them.

Ejection of gas is observed as galactic outflows driven by massive stars, supernovae, or active galactic nuclei (AGN).  This has been proposed to explain the origin of \foiii+\hb\ halos and CGM enrichment in galaxies in the distant universe \cite{Parlanti2024,Marshall2023,Jones2024}. The large \foiii+\hb\ halos of the size seen in JQ have only been observed in the quasars and AGNs \cite{Marshall2023,Vayner2023,Solimano2024}. 
There is currently no evidence of AGN in the JQ galaxies, particularly in ELG1 and ELG2 given the spatially resolved morphology of their clumps. 
There is also no evidence of any broad emission component in the existing NIRSpec M-grating spectrum of ELG1, indicating if such an outflow exists, it is not powerful enough to eject large amounts of oxygen-enriched gas to CGM, at least along the line of sight.
We compared the properties of outflows in a state-of-the-art model and found the outflows produced by supernovae in ELG1 and ELG2 and their adjacent clumps are not sufficient to explain the observed luminosity of the \foiii+\hb\ ``bridge'' between the two galaxies (see Methods).
Consequently, the outflows are unlikely to be the dominant driver of CGM enrichment in JQ.

It is therefore more plausible that the \foiii+\hb\ halo of JQ results from oxygen-enriched gas stripped out of the galaxies through interactions and tidal forces \cite{DiCesare2024,Sparre2022}.  
The stripped gases are shock-heated in galaxy collisions \cite{Moreno2019,Sinha2009}, and dispersed to large radii, leading to large hot gaseous halos \cite{Cox2006}.
This scenario is believed to be responsible for the diffuse emission halo in Stephan's Quintet, where the shocked-heated gas has a temperature of $T>6\times10^6$ K and appears as a ridge of X-ray and radio synchrotron emission \cite{O'Sullivan2009,Allen1972}, residing between the two colliding galaxies.
Such emission is below the sensitivity of X-ray and radio telescopes given JQ's distance, but it is possible to confirm the shock-heated origin of the gas using excitation diagnostics of emission lines, as those have been performed in Stephan's Quintet \cite{Xu2003,Iglesias-Paramo2012,Rodriguez-Baras2014}.
Similar to the distribution of shock-heated gas in Stephan's Quintet, the majority of diffuse \foiii+\hb\ emission in JQ also arises from the ``bridge'' between two merging galaxies, ELG1 and ELG2.
Therefore, the evidence supports the interpretation that the \foiii+\hb\ halo in JQ results from shock-heated gas, stripped from the galaxies through interactions from a merger. 

The discovery of the JQ system at $z=6.71$ in the JADES data suggests that major mergers of multiple galaxies might be more frequent in the early universe than expected from theory. 
These mergers can efficiently eject metal-enriched gas into the CGM, thereby contributing to the development of the galactic ecosystem.  The elevated star formation associated with mergers offers a unique solution to the tension between the observation and cosmological simulation on the observed overabundance of massive galaxies and massive quiescent galaxies recently unveiled by JWST.

\clearpage
\section{Methods}

Throughout this study, we adopt the Planck 2018 cosmological parameters \cite{PlanckCollaboration2020}: $\Omega_m=0.3111$, $\Omega_\Lambda=0.6889$ and $H_0=67.66$ \kms\ Mpc$^{-1}$, where $\Omega_m$ and $\Omega_\Lambda$ are the densities of total matter and dark energy and $H_0$ is the Hubble constant.

\editone{
We acknowledge that in this work, we use the term ``clump'' to describe a compact, spatially unresolved structure in the JWST image. 
We define a galaxy as an isolated system composed of one or more clumps with similar stellar populations; we do not impose a specific minimum number of clumps for this definition.
An emission line galaxy is defined as a galaxy showing strong equivalent widths in nebular lines, such as \foiii+\hb.}

\subsection{HST and JWST Imaging data}
We utilize the published JWST images and spectroscopy from the JWST Advanced Deep Extragalactic Survey (JADES) DR3 \cite{Eisenstein2023,Eisenstein2023b,D'Eugenio2024} to measure the flux densities and to detect clumps in the JQ galaxies.
JADES is one of the deepest NIRCam surveys, providing 14 broad- or medium-bands (broad- and medium-band imaging from F090W, F115W, F150W, F182M, etc. filters from JADES \cite{Hainline2024}, combined with medium-band imaging in F430M, F460M, and F480M from JEMS \cite{Williams2023}) covering wavelengths from 0.8 to 5.0 $\mu$m.
The JADES photometric catalog provides matched-aperture flux densities for the galaxies, and photometric redshifts measured from the SEDs \cite{Rieke2023,Eisenstein2023b,Hainline2024}.

In addition to the JWST images, we also utilize the archival Hubble Space Telescope (HST) images to complement the JWST images in the UV-to-optical wavelength range and to remove the foreground galaxies.
The archival HST images were obtained from several HST observation programs, including CANDELS, HUDF, UVCANDELS, etc \cite{Grogin2011,Koekemoer2011,Wang2025,Bouwens2011}.
We adopt the reduced HST images provided by the JWST FRESCO collaboration \cite{Oesch2023}, which has already been registered to match the JWST pixel scale (0.03 arcsec).

\subsection{JWST Spectroscopy data}
The spectroscopic data of ELG1 and ELG5 used in this work were also obtained as part of the JADES NIRSpec observations.
The JADES program opened three adjacent shutters for each target and nodded between these shutters (perpendicular to the dispersion direction) to improve the background subtraction and bad pixel rejection.
JADES obtained low-resolution spectra ($R\sim100$) using the NIRSpec disperser/filter configuration of PRISM/CLEAR, which provides wavelength coverage from 0.6 to 5.3 $\mu$m.
The exposure times of ELG1 and ELG5 were $\sim$ 2.1 and 9.3 hours, respectively.
JADES also obtained medium-resolution spectra ($R\sim1000$) using the disperser/filter configurations of G140M/F100LP, G140M/F070LP, and G235M/F170LP, G395M/F290LP, which provides a full wavelength coverage from 0.7 to 5.3 $\mu$m.
The exposure times of ELG1 for the three configurations were 3.4, 2.8, and 2.8 hours, and the exposure times of ELG5 for the three configurations were 2.3 hours.
The JADES NIRSpec data used in this work were downloaded from the MAST HLSP archive\footnote{\url{https://archive.stsci.edu/hlsp/jades}}. 
Please refer to ref \cite{Bunker2024} for details.

\subsection{Photometric redshifts of ELGs}

Based on the JADES photometry, we use the Python package \texttt{Bagpipes} \cite{Carnall2018} to perform SED fitting, including measuring redshifts of ELG2, ELG3, and ELG4.
\editone{We note that although the JWST images show ELG3 and ELG4 likely connected, we consider them as two separate galaxies because of their distinct stellar populations. 
As shown in Figure \ref{fig:mass_sfr_galaxy} and Extended Data Table \ref{tab:properties}, ELG3 is the most massive galaxy with a moderate SFR consistent with being located on the SFMS. In contrast, ELG4 has significantly elevated SFR, placing it above SFMS by $>1$ dex. This indicates ELG3 is dominated by more mature populations, but ELG4 is dominated by newly born stars.}
We employ Binary Population and Spectral Synthesis v2.2.1 \cite{Eldridge2017,Stanway2018} (BPASS) with a broken power-law initial mass function with slopes of $\alpha_1 = -1.3$ for stars with 0.1 -- 0.5 $M_\odot$ and $\alpha_2 = -2.35$ for 0.5 -- 100 $M_\odot$ (model `135\_100').
For the star formation history parameterization, we use the Gaussian Process model from \texttt{dense\_basis} \cite{Iyer2019}, where the star formation history is split into 4 dynamically adjusted time bins, and during each time bin, $25\%$ of the total stellar mass is formed.
The metallicity is allowed to span 0.001 -- 1~$Z_\odot$.  
We adopt the Calzetti \cite{Calzetti2000} dust attenuation law with $A(V)$ ranging from 0.0 to 4.0~mag.  
We include nebular emission with the metallicity of the gas equal to that of the stellar populations, and an ionization parameter,  $\log U$ in the range $-4$ to $-1$. 
We use the uniform priors for the \texttt{Bagpipes} parameters across the allowed range.
For the redshift, we also adopt a uniform prior across the range of \editone{0} -- 8.5.
Figure \ref{fig:photoz} shows the probability distribution function of the redshifts and the best-fit models for ELG2, ELG3, and ELG4.

We also use \texttt{Bagpipes} to estimate the slitlosses of NIRSpec spectroscopy of ELG1 and ELG5.
We adopt a multiplicative scaling factor (assumed to be a second-order Chebyshev polynomial) in \texttt{Bagpipes} to correct the wavelength-dependent spectral flux calibration to the broadband photometry.
We allow the zeroth order to vary from 0.1 -- 10 because of the relatively large slitloss in ELG1, while the first and second orders are allowed to vary from -0.5 -- 0.5. 
In Figure \ref{fig:photoz}, we present the flux-calibrated spectra of ELG1 and ELG5.

\addtocounter{figure}{-4} 
\begin{figure*}
    \renewcommand{\figurename}{Extended Data Figure}
    \centering
    \includegraphics[width=0.8\linewidth]{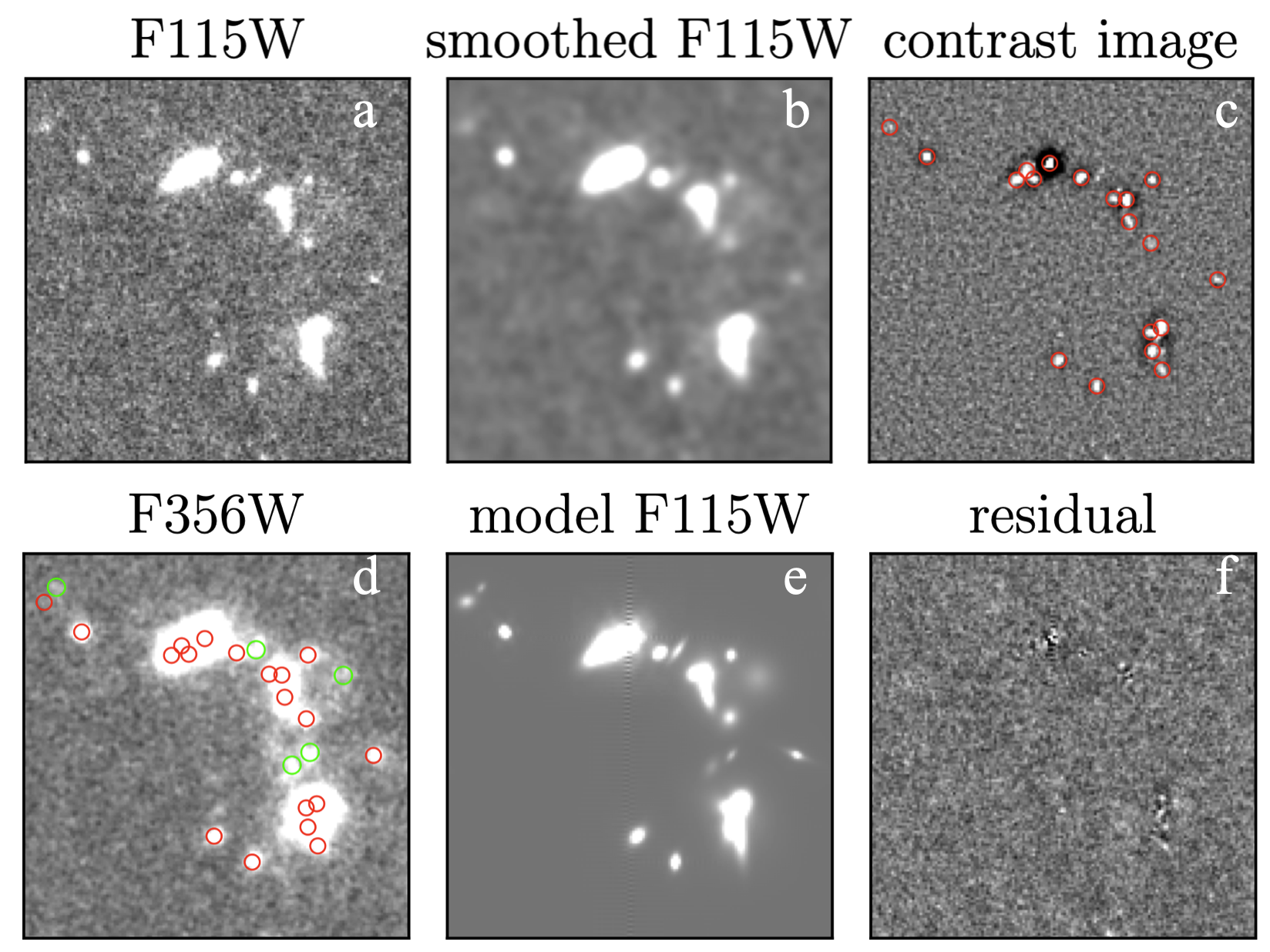}
    \caption{\textbf{Illustration of clump identification.} \textbf{a)~} The original F115W image of JQ. We adopt the F115W as the detection image because it has the best spatial resolution and is not affected by the \lya\ breaks of the galaxies.
    \textbf{b)~} The smoothed F115W image of JQ. We smooth the original F115W image by a two-dimensional Gaussian Kernel with $\sigma=2$ pixel (0.06 arcsec).
    \textbf{c)~} The contrast image, which is derived by subtracting the smoothed F115W image from the original F115W image. This step removes the outshining component and enhances the contrast of clump cores. We detect the clumps from this contrast image and mark the automatically-identified clumps as red circles.
    \textbf{d)~} The F356W image of JQ. We notice several clumps that are only marginally detected in F115W but are clearly visible in F356W (green circles). Therefore, we manually add them to the catalog.
    \textbf{e)~} The best-fit S\'ersic models of all galaxies in the image. This includes all galaxies and clumps.
    \textbf{f)~} The residual map after subtracting the best-fit S\'ersic models for galaxies and the clumps from the original F115W image. The lack of large structures indicates we have well-modeled the light in these galaxies and their clumps.}
    \label{fig:clumps}
\end{figure*}

\begin{sidewaystable}
    \renewcommand{\tablename}{Extended Data Table}
    \renewcommand{\arraystretch}{1.5}
    \centering
    \begin{tabular}{c c c c c c c c c c c}
    \hline
    \hline
        Galaxy & clump & RA & DEC & $\log\ M_\star$ & SFR$_\mathrm{10Myr}$  & M$_\mathrm{UV}$ & $A_V$ & $\log\ L_\mathrm{[\mathrm{O~\textsc{III}}]+\mathrm{H}\beta}$ & $r_\mathrm{eff}$ \\
        ID & ID &  &  & $\log\ [M_\odot]$ & $M_\odot$ yr$^{-1}$ &  & & $\log$ [erg s$^{-1}$] & pkpc \\
        \hline
        \multirow{3}{*}{ELG1} & C1 & 3:32:37.19 & -27:48:57.04 & $8.12^{+0.02}_{-0.02}$ & $12.9^{+0.5}_{-0.4}$ & $-19.66^{+0.03}_{-0.03}$ & $0.46^{+0.02}_{-0.02}$ & $42.72{\pm}0.06$ & $0.45{\pm}0.06$ \\
        & C2 & 3:32:37.19 & -27:48:56.83 & $8.72^{+0.06}_{-0.04}$ & $47.5^{+2.4}_{-2.3}$ & $-19.37^{+0.03}_{-0.03}$ & $1.12^{+0.03}_{-0.03}$ & $42.90{\pm}0.08$ & $0.65{\pm}0.12$ \\
        & C3 & 3:32:37.18 & -27:48:56.76 & $9.19^{+0.03}_{-0.07}$ & $9.1^{+0.9}_{-0.7}$ & $-19.28^{+0.04}_{-0.04}$ & $0.57^{+0.07}_{-0.05}$ & $42.47{\pm}0.06$ & $0.40{\pm}0.03$ \\
        \hline
        & C4 & 3:32:37.19 & -27:48:55.75 & $8.41^{+0.16}_{-0.19}$ & $9.3^{+2.2}_{-2.0}$ & $-17.87^{+0.12}_{-0.14}$ & $1.07^{+0.11}_{-0.12}$ & $42.12{\pm}0.12$ & $0.67{\pm}0.17$ \\
        \hline
        \multirow{2}{*}{ELG2} & C5 & 3:32:37.21 & -27:48:55.50 & $8.00^{+0.06}_{-0.05}$ & $8.8^{+0.5}_{-0.5}$ & $-18.60^{+0.06}_{-0.06}$ & $0.72^{+0.05}_{-0.05}$ & $42.40{\pm}0.06$ & $0.65{\pm}0.08$ \\
        & C6 & 3:32:37.21 & -27:48:55.26 & $8.00^{+0.10}_{-0.07}$ & $8.2^{+0.3}_{-0.3}$ & $-19.85^{+0.03}_{-0.03}$ & $0.21^{+0.02}_{-0.02}$ & $42.59{\pm}0.06$ & $0.39{\pm}0.05$ \\
        \hline
        & C7 & 3:32:37.19 & -27:48:55.03 & $7.30^{+0.13}_{-0.07}$ & $1.6^{+0.2}_{-0.1}$ & $-17.74^{+0.10}_{-0.09}$ & $0.33^{+0.08}_{-0.08}$ & $41.87{\pm}0.06$ & $0.12{\pm}0.06$ \\
        \hline
        \multirow{3}{*}{ELG3} & C8-1 & 3:32:37.30 & -27:48:54.92 & \multirow{2}{*}{$9.88^{+0.04}_{-0.06}$} & \multirow{2}{*}{$12.9^{+1.9}_{-1.7}$} & \multirow{2}{*}{$-19.87^{+0.03}_{-0.03}$} & \multirow{2}{*}{$0.68^{+0.08}_{-0.10}$} & \multirow{2}{*}{$42.29{\pm}0.12$} & $0.56{\pm}0.02$ \\
        & C8-2 & 3:32:37.30 & -27:48:55.02 &  & & & & & $0.13\pm0.03$ \\
        & C9 & 3:32:37.31 & -27:48:55.03 & $8.92^{+0.04}_{-0.03}$ & $2.0^{+0.7}_{-0.5}$ & $-19.42^{+0.05}_{-0.05}$ & $0.21^{+0.08}_{-0.10}$ & $41.46{\pm}0.40$ & $0.55{\pm}0.04$ \\
        \hline
        ELG4 & C10 & 3:32:37.28 & -27:48:54.83 & $8.44^{+0.02}_{-0.03}$ & $27.3^{+0.1}_{-0.2}$ & $-21.38^{+0.01}_{-0.01}$ & $0.12^{+0.01}_{-0.01}$ & $43.21{\pm}0.10$ & $0.09{\pm}0.01$ \\
        \hline
        ELG5 & C11 & 3:32:37.39 & -27:48:54.75 & $7.50^{+0.04}_{-0.03}$ & $2.9^{+0.1}_{-0.2}$ & $-18.29^{+0.04}_{-0.05}$ & $0.37^{+0.04}_{-0.04}$ & $42.20{\pm}0.06$ & $0.09{\pm}0.01$ \\
        \hline
        & C12 & 3:32:37.19 & -27:48:56.20 & $7.54^{+0.13}_{-0.10}$ & $2.6^{+0.6}_{-0.4}$ & $-16.81^{+0.21}_{-0.20}$ & $0.90^{+0.13}_{-0.13}$ & $41.82{\pm}0.12$ & $0.35{\pm}0.22$ \\
        & C13 & 3:32:37.21 & -27:48:56.33 & $7.60^{+0.14}_{-0.11}$ & $3.0^{+0.7}_{-0.5}$ & $-16.51^{+0.23}_{-0.20}$ & $1.09^{+0.17}_{-0.16}$ & $41.85{\pm}0.12$ & $0.47{\pm}0.22$ \\
        & C14 & 3:32:37.17 & -27:48:55.28 & $7.98^{+0.27}_{-0.26}$ & $3.0^{+0.7}_{-0.8}$ & $-18.44^{+0.16}_{-0.13}$ & $0.39^{+0.13}_{-0.14}$ & $41.76\pm0.20$ & $1.00{\pm}0.12$ \\
        & C15 & 3:32:37.41 & -27:48:54.24 & $7.05^{+0.43}_{-0.34}$ & $0.6^{+0.4}_{-0.3}$ & $-16.31^{+0.43}_{-0.31}$ & $0.56^{+0.31}_{-0.25}$ & $41.28\pm0.36$ & $0.37{\pm}0.26$ \\
        & C16 & 3:32:37.13 & -27:48:56.19 & $8.91^{+0.07}_{-0.06}$ & $0.005^{+0.057}_{-0.005}$ & $-17.31^{+0.14}_{-0.13}$ & $0.29^{+0.21}_{-0.16}$ & -- & $0.46{\pm}0.18$ \\
        \hline
        JQ & & & & $9.99^{+0.03}_{-0.02}$ & $255.2^{+12.9}_{-14.2}$ & & & $43.77{\pm}0.10$ & \\
    \hline
    \end{tabular}
    \caption{\textbf{Physical properties of the 17 galaxy-size clumps and the entire JQ system.} The physical properties of clumps are obtained from fitting the SEDs of individual clumps, while the total stellar mass and SFR of JQ are determined through a spatially resolved analysis of the entire JQ system. The total \foiii+\hb\ luminosity is derived by summing the \foiii+\hb\ map. See Methods for details about the SED fitting.}
    \label{tab:properties}
\end{sidewaystable}

\subsection{Clump identification and photometry} \label{sec:clump}

Our clump identification method is illustrated in the Extended Data Figure \ref{fig:clumps}.
We identify the clumps from the NIRCam F115W image, as it offers the highest spatial resolution compared to the NIRCam images at longer wavelengths and is unaffected by the \lya\ break,  unlike F090W.
We first extract a $4.5''\times4.5''$ cutout, corresponding to a physical size of $\sim 25\times25$ pkpc$^{2}$.
To suppress the outshining components and enhance the visibility of clump cores, we smooth the cutout using a two-dimensional Gaussian kernel and subtract the smoothed image from the original F115W cutout to produce a contrast image. 
\editone{A kernel that is too narrow (e.g., $\sigma=1$ pixel) excessively subtracts flux from the clump cores, making them difficult to detect, while a kernel that is too broad (e.g., $\sigma=4$ pixel) can produce a large dark halo surrounding the central bright clump, resulting in contamination to its nearby objects.
To balance between these two effects, we adopt a kernel size of $\sigma=2$ pixel (0.06 arcsec), of which the FWHM is comparable to the separation between close clumps in JQ. This configuration effectively preserves the clump cores while minimizing contamination to neighboring sources.}
We then use the Python Package \texttt{sep} \cite{Barbary2016, Bertin1996} to detect clumps in the contrast image and extract their positions.
The red circles in Extended Data Figure \ref{fig:clumps}c,d mark the \editone{automatically} identified clumps.
\editone{However, as F115W traces the rest-frame UV emission of objects, which is sensitive to dust attenuation, this method could miss clumps that are heavily obscured.
In contrast, the F356W image probes the rest-frame optical emission of objects, which is less affected by dust and can be significantly enhanced by strong \foiii+\hb\ emission. 
Therefore, F356W provides a valuable complement for identifying clumps that are attenuated.
Through a visual inspection of the F356W image, we identify 5 additional clumps that are missed in the F115W contrast image.}
These clumps are then manually added to the catalog based on their positions in the F356W image, and they are marked as green circles in Extended Data Figure \ref{fig:clumps}d.
\editone{In total, we identify 17 clumps in JQ: 3 in ELG1, 2 in ELG2, 3 in ELG3, 1 each in ELG4 and ELG5, and 7 additional clumps in the surrounding region.}
\editone{We acknowledge that the spatially inhomogeneous dust distribution could artificially divide a clump into multiple clumps, leading to an overestimation of the number of clumps.
Confirming this would require a high-resolution, dust-insensitive observation, such as ALMA [C II] observations.
Nonetheless, this does not affect our subsequent comparison with simulations, because we consider the close clumps as a single system in the analysis.}

We measure the clump photometry by modeling each clump with a S\'ersic profile convolved with the corresponding point spread function (PSF) for each band using the Python package \texttt{Pysersic} \cite{Pasha2023}. 
\texttt{Pysersic} requires a set of priors, including clump positions, fluxes, and S\'ersic parameters to initialize the fitting.
It applies the No-U Turn (NUTS) sampler \cite{Hoffman2011} to simultaneously fit all clumps and produce the posterior distributions.
The PSF used in \texttt{Pysersic} is measured using the compact, isolated stars identified in the JADES NIRCam images in each filter.

We first perform the morphological fitting to the F115W image. 
We adopt the uniform priors for the S\'ersic parameters and the Gaussian priors for the positions with the mean to be those measured by \texttt{sep}.
\editone{For the fluxes, we adopted the Gaussian priors with the mean of 0 $\mu$Jy and the standard deviation of 0.2 $\mu$Jy to cover all the possible ranges.}
Extended Data Figure \ref{fig:clumps} panels $e$ and $f$ show the best-fit model and the residual map of F115W, respectively.
\editone{The residual map of F115W does not show clear large structure, indicating that we have identified all the bright components and well-modeled their lights.}
After obtaining the best-fit parameters from F115W, we use these measurements as the priors for fitting the clumps in other bands.
Due to the substantial PSF variation across JWST images (from $0.04''$ in F115W to $0.15''$ in F444W), deblending adjacent clumps becomes increasingly challenging at longer wavelengths.
Therefore, to address this, we fix the $x,y$ pixel centers and S\'ersic parameters to the values obtained from F115W, and only the fluxes are allowed to vary when fitting the other bands (F090W, F150W, F200W, F277W, F335M, F356W, F410M, and F444W).

\subsection{Photometric redshifts and physical properties of clumps} \label{sec:clumpp}
Using the photometry described in the previous section, we perform model fitting for the clumps again using \texttt{Bagpipes}.
This step is necessary to identify the clumps that are associated with JQ and remove other foreground clumps unassociated with the JQ galaxies.
The \texttt{Bagpipes} configuration is identical to the one used for the photometric redshift estimation of ELG2, ELG3, and ELG4.
We visually inspect the best-fit results and define a clump as a member of JQ if its best-fit photometric redshift lies within the range of 6.65 and 6.75. 
Once the member clumps of JQ are identified, we rerun the SED fitting with the redshift fixed to 6.71 to best model their properties assuming the spectroscopic redshift of ELG1 and ELG5.
We note that clumps C8-1 and C8-2 are heavily blended in the longer wavelength imaging data, so we combine their photometry for the SED fitting process. 
In Extended Data Table \ref{tab:properties}, we summarize the physical properties of the identified member clumps.

\begin{figure}
    \renewcommand{\figurename}{Extended Data Figure}
    \centering
    \includegraphics[width=0.8\linewidth]{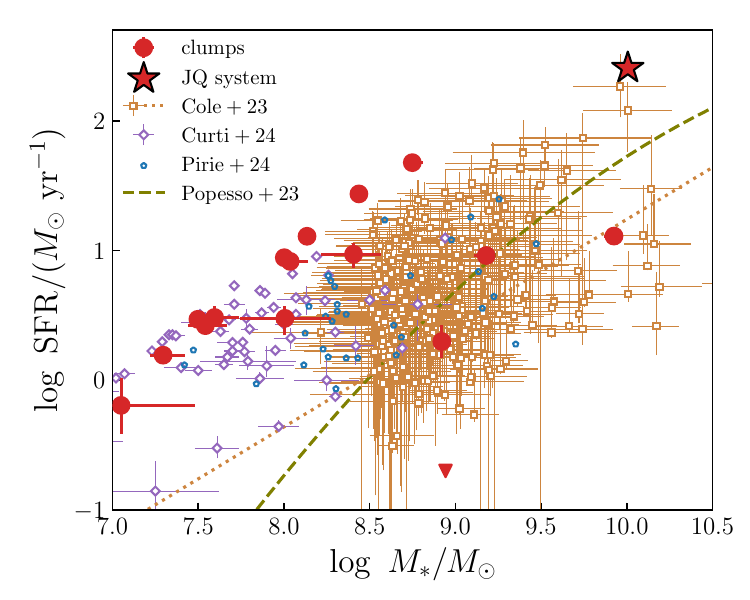}
    \caption{\textbf{Similar to Figure \ref{fig:mass_sfr_galaxy}, but with the five galaxies replaced by their member clumps.} }
    \label{fig:mass_sfr}
\end{figure}

\editone{
In Figure \ref{fig:mass_sfr_galaxy} and Extended Data Figure \ref{fig:mass_sfr}, we compare the SFRs of the JQ galaxies and clumps with the galaxies and the SFMS at similar redshift from the literature \cite{Cole2023,Curti2024,Pirie2024,Popesso2023}.
Most of the galaxies and clumps in JQ show elevated SFRs compared to the median SFMS relation. 
Recently, it has been recognized that the high-redshift galaxies experience bursty star formation \cite{Endsley2024}, where a low-mass galaxy in the bursty phase can significantly increase its opportunity of being detected. 
However, this observational bias is unlikely to affect our analysis of the JQ clumps. 
JQ is in the deepest region of the JADES survey, which provides sufficient sensitivity to detect a complete low-mass sample reaching $<10^8\ M_\odot$. 
Specifically, the purple diamonds in Extended Data Figure \ref{fig:mass_sfr} are the spectroscopically observed galaxies selected from the JADES survey \cite{Curti2024}.
This sample includes a large number of galaxies with stellar masses comparable to those of the JQ clumps, but with much lower SFRs. 
Therefore, the elevated SFRs observed in the JQ clumps are intrinsic rather than due to this selection effects.}

\begin{figure*}
    \renewcommand{\figurename}{Extended Data Figure}
    \centering
    \includegraphics[width=0.48\linewidth]{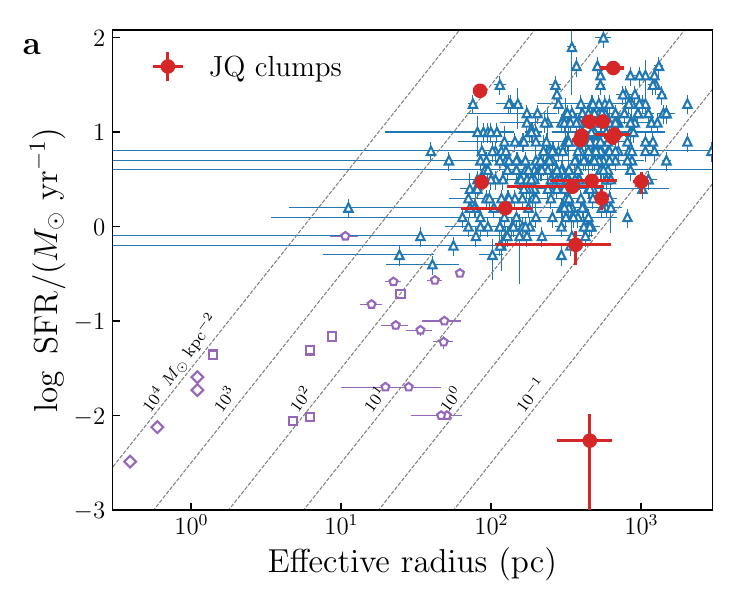}
    \includegraphics[width=0.48\linewidth]{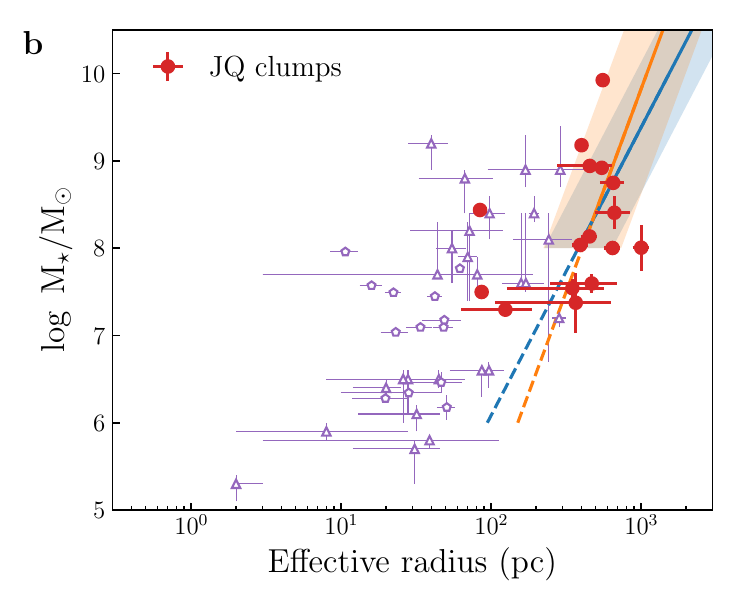}
    \caption{\textbf{Size -- SFR relation and size -- mass relation of the JQ clumps.} \textbf{a)} The size -- SFR relation. The red dots indicate the individual clumps in the JQ system. For comparison, we compile several objects from the lensing galaxies and field galaxies at similar redshifts. The blue open triangles represent the field galaxies at $z\sim5$ -- 14 \cite{Morishita2024}. The purple open diamonds, squares, and pentagons represent the star clusters and star-forming clumps identified within the individual lensing galaxies at $z>6$ \cite{Vanzella2023,Adamo2024,Fujimoto2024}. The dashed lines indicate the star formation rate surface densities of $10^4$, $10^3$, $10^2$, $10^1$, $10^0$, and $10^{-1}$ $M_\odot$ kpc$^{-2}$ from left to right. \textbf{b)} \editone{The size -- mass relation. We present the JQ clumps as the red dots. The purple triangles and pentagons represent the star clusters and star-forming clumps identified within the individual lensing galaxies at $z>5$ \cite{Fujimoto2024,Claeyssens2023}. The blue and orange solid lines indicates the best-fit size -- mass relation for $5\leqslant z < 6 $ and $6\leqslant z<9$ galaxies \cite{Allen2024}, and the shaded regions represent their $1\sigma$ scatter. The dashed lines represent the extrapolation of these relations.}}
    \label{fig:sizesfr}
\end{figure*}

In Extended Data Figure \ref{fig:sizesfr}, we compare the size -- SFR relation \editone{and size -- mass relation} of the clumps in the JQ system with those of the galaxies at similar redshifts observed with JWST NIRCam \cite{Vanzella2023,Adamo2024,Fujimoto2024,Morishita2024,Claeyssens2023,Allen2024}. 
Several studies \cite{Vanzella2023,Adamo2024,Fujimoto2024,Claeyssens2023} resolved the $z>5$ galaxies through the lensing technique and found that the star clusters and star-forming clumps identified within the individual galaxies have sizes $<100$ pc and SFRs $<1\ M_\odot$ yr$^{-1}$. 
Their sizes and SFRs are significantly smaller than those of the clumps in the JQ system. 
The JQ clumps have sizes and SFRs more similar to those unresolved galaxies at $z\sim$ 5 -- 14 \cite{Morishita2024}, which typically have sizes $>100$ pc and SFRs $>0.3\ M_\odot$ yr$^{-1}$.
\editone{In addition, as shown in Extended Data Figure \ref{fig:sizesfr}, the JQ clumps generally follow the size -- mass relation measured for $z>5$ galaxies, whereas the lensing-resolved clumps in previous studies have much smaller sizes at a fixed mass. 
Therefore, to distinguish them from these smaller structures, we refer to the JQ clumps as galaxy-size clumps.
Given the limited resolution of JWST observations, we do not attempt to further classify the clumps into morphological components such as bulges and disks in this work.}

\subsection{\foiii+\hb\ emission map of JQ} 

\begin{figure}
    \renewcommand{\figurename}{Extended Data Figure}
    \centering
    \includegraphics[width=\linewidth]{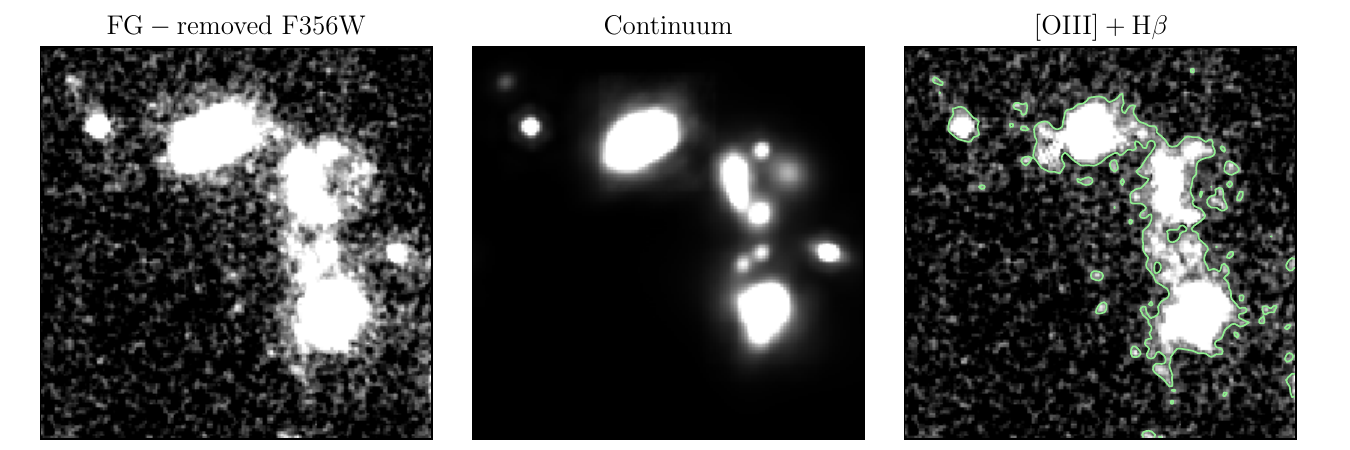}
    \caption{\textbf{Illustration of generating the \foiii+\hb\ map.} \textbf{a)~} The F356W image after removing the foreground galaxies.
    \textbf{b)~} The continuum model generated based on the best-fit SEDs of clumps, the S\'ersic profiles of clumps, and the F356W PSF. 
    \textbf{c)~} The \foiii+\hb\ map derived by subtracting the continuum model from the foreground-removed F356W image. The green contour indicates surface brightnesses of 8.8 $\times10^{-17}$ erg s$^{-1}$ cm$^{-2}$ arcsec$^{-2}$, the same as Figure \ref{fig:rgb}.}
    \label{fig:oiiimap}
\end{figure}

The \foiii+\hb\ emission map of JQ is created by subtracting the continuum contribution from the NIRCam F356W image. 
Previous works often use images at slightly longer or shorter wavelengths to estimate the continuum contribution, such as NIRCam F335M for our case.
However, due to the NIRCam F335M image being much shallower than the F356W, subtracting F335M would significantly reduce the SNR, making the diffuse \foiii+\hb\ halo invisible.

In this work, we adopt an alternative method to estimate the continuum contribution.
First, we remove the foreground galaxies using their best-fit flux in F356W and S\'ersic models. 
We then estimate the continuum for clumps associated with JQ based on their best-fit spectra from the \texttt{Bagpipes} fitting.
We mask the \foiii\ $\lambda\lambda4959,5007$ and \hb\ $\lambda4861$ emission lines in the best-fit spectrum and convolve them with the transmission curve of the F356W filter to calculate their contribution to the F356W image.
We then use their best-fit S\'ersic models, the continuum flux density, and the F356W PSF to generate a continuum map (shown in Extended Data Figure \ref{fig:oiiimap}b).
We subtract this continuum map from the foreground-removed F356W image to generate the \foiii+\hb\ emission map, as shown in Extended Data Figure \ref{fig:oiiimap}c.

\subsection{Spatially-resolved photometry and properties}
As aforementioned, we fix the half-light radii of clumps when fitting their morphologies in the F356W image. 
However, emission lines are frequently observed to be more spatially extended than the underlying stellar continuum \cite{Shen2023}, so this approach could miss diffuse \foiii+\hb\ emission from the clumps, leading to an underestimation of their flux densities.
The simple S\'ersic profile is not always able to accurately describe the morphologies of the clumps due to their disturbed and irregular shapes. 
\editone{Moreover, when galaxies show high specific star formation, the young, luminous stellar populations can outshine the older populations and dominate the light, resulting in further underestimation of stellar mass \cite{Sorba2018}.}
To address these limitations and provide a robust estimate of the total stellar mass and SFR of the JQ system, we perform spatially resolved SED fitting directly using multi-wavelength images.
This method is insensitive to the assumption of morphologies across different wavelengths and has been extensively validated in previous studies \cite{Shen2024}.
For this analysis, we adopt the SED fitting package, \texttt{CIGALE} \cite{Boquien2019}, which offers computational efficiency when fitting a large number of data points. 
This also allows us to assess the potential biases associated with the SED fitting methods. 

\begin{figure}
    \renewcommand{\figurename}{Extended Data Figure}
    \centering
    \includegraphics[width=\linewidth]{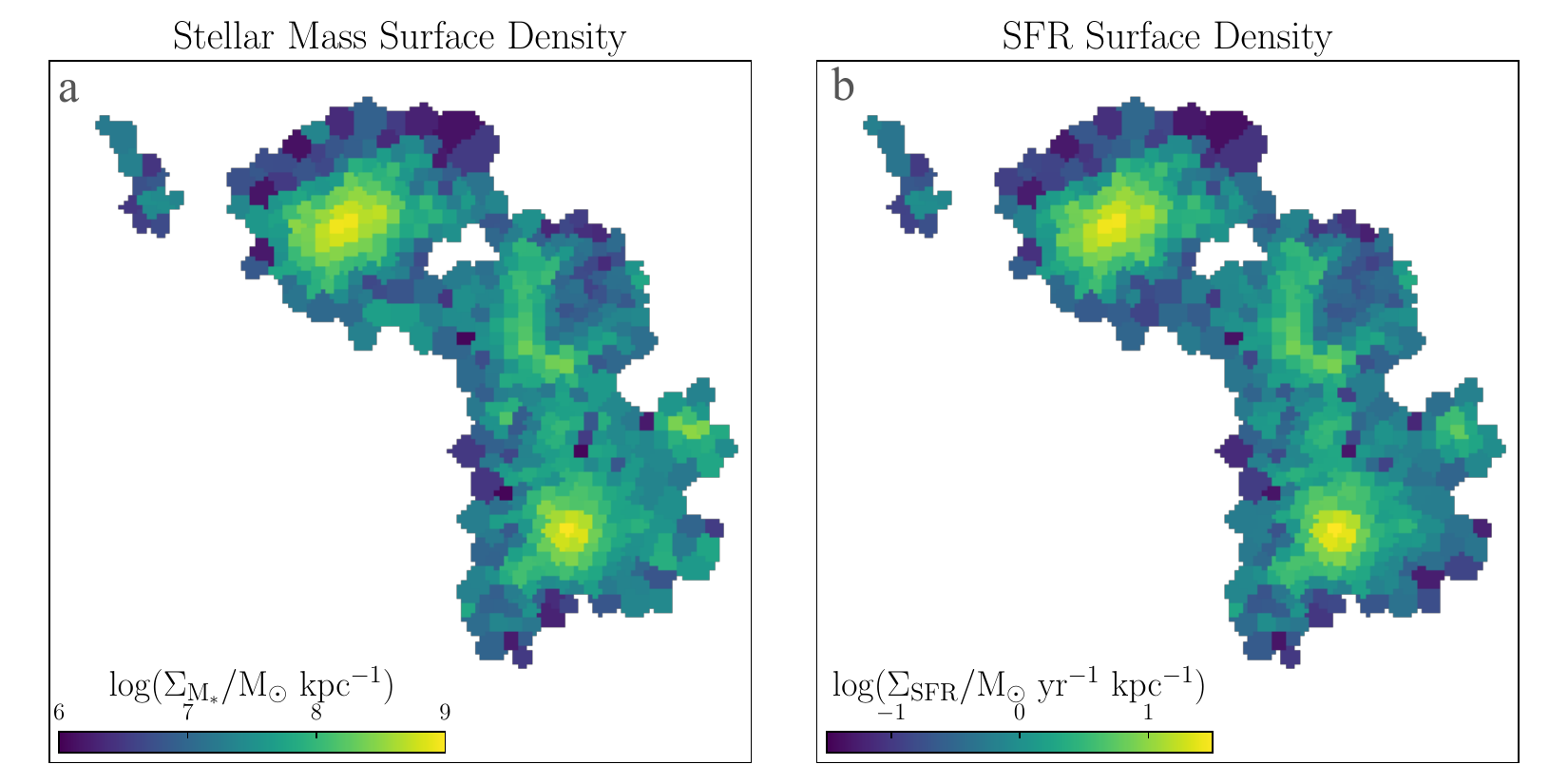}
    \caption{\textbf{Spatially resolved maps of mass and SFR.} \textbf{a,} The stellar mass surface density map. \textbf{b,} The SFR surface density map.}
    \label{fig:cigale}
\end{figure}

Our method follows the procedure in ref \cite{Shen2024}.
In detail, we used the PSF-matched images \editone{(to F444W)} with a pixel scale of 0.03 arcsec. 
To enhance the SNR of regions surrounding and connecting the galaxies in the JQ, we applied an adaptive binning algorithm based on the Weighted Voronoi Tessellation method described in ref \cite{Cappellari2003, Diehl2006}. 
We chose a minimum SNR of 8 in the F356W band, and a minimum Voronoi bin size of 9 pixels, which is approximately the size of the FWHM of the PSF of F356W (FWHM = 0\farcs1). 
\editone{
It has been recognized that, with increasing spatial resolution, the smallest resolved elements may have too little stellar mass for the stellar initial mass function (IMF) to be fully populated.
This effect, known as stochastic IMF sampling, can bias the derived properties such as stellar mass, age, and SFR. 
Our binning procedure ensures that the typical stellar masses of the resulting bins are $>10^{5.4}\ M_\odot$, which is well above the threshold ($\sim10^5\ M_\odot$; e.g., \cite{Applebaum2020}) where stochastic IMF sampling is expected to significantly affect results. Therefore, this effect does not impact our analysis.}

We run \texttt{CIGALE} using the 9-band fluxes in each Voronoi bin. 
Here we describe the modules and parameters adopted in \texttt{CIGALE}. We adopt a delayed SFH allowing the $\tau$ and stellar age to vary from 0.05 -- 1~Gyr and 40 -- 800~Myr, respectively. 
In addition to the main SFH, we allow for a more recent  ``burst'' of star formation as an exponential SFH with a $\tau$ allowed to vary from 20 -- 100~Myr, an age of 1 -- 30 Myr, and a mass fraction of this burst ($f_\mathrm{burst}$) to vary from 0 -- 0.9. 
We assume a ref \cite{Chabrier2003} IMF and the stellar population synthesis models presented by ref \cite{Bruzual2003} with metallicity ranging from Solar ($Z_\odot$) to sub-Solar values (0.4~$Z_\odot$ and 0.2~$Z_\odot$). 
We include nebular emission using templates of ref \cite{Inoue2011}. We allow the ionization parameter $\log(U)$ to vary between $-3$ to $-1$, the gas metallicity ($Z_\mathrm{gas}$) to vary between 0.002 and 0.02, and a fixed electron density of 100~cm$^{-3}$. 
The dust attenuation follows the Calzetti extinction law \cite{Calzetti2000} allowing the dust attenuation in emission lines from nebular regions $E(B-V)_l$ to vary from 0 to 1.1, and a lower dust attenuation in stellar continuum with a fixed dust attenuation ratio ($E(B-V)_\mathrm{star} / E(B-V)_l = 0.44$) between emission lines and stellar continuum. 
In Extended Data Figure \ref{fig:cigale}, we present the spatially resolved maps of mass and SFR. 

The integrated properties of the JQ system are derived by summing the spatially resolved maps.
We obtain a total stellar mass of $\log(M_*/M_\odot) = 9.99^{+0.03}_{-0.02}$ and a total SFR averaged of $255.2^{+12.9}_{-14.2}$~$M_\odot$~yr$^{-1}$. 
We estimate the uncertainties in the total stellar mass and SFR by propagating the measurement errors from each Voronoi bin.
The total stellar mass is consistent with the sum of the stellar masses of individual clumps from \texttt{Bagpipes}.
\editone{This consistency implies that we do not identify the stellar population variation at a clump-size scale, and the majority of stellar mass is formed within the clumps.}
However, the total SFR is nearly double the sum of the SFRs of the clumps (see Extended Data Table \ref{tab:properties}). 
To investigate this discrepancy, we use \texttt{CIGALE} to model the SEDs of 17 clumps and find that the estimated stellar masses and SFRs agree with those derived from \texttt{Bagpipes} within $1\sigma$ uncertainties. 
Therefore, the difference between the SFRs estimated from the spatially resolved map and the individual clumps is not due to the choice of the SED fitting method.
\editone{Instead, the difference in the derived SFRs suggests that a large fraction, $\leq40\%$, of the total SFR may arise from diffuse emission surrounding the clumps if attributing the diffuse emission to star formation. }

\subsection{Contextualizing JQ-like systems with cosmological simulations}

We compare the presence and properties of the JQ system with galaxy-size clumps found in a simulated lightcone \cite{Yung2023}.
The simulated lightcone spans 2 deg$^2$, containing simulated galaxies arranged in a way that mimics how we observe the real Universe. The dataset provides galaxy coordinates (RA and Dec) on the projected plane, and includes both cosmological redshifts and peculiar velocities of galaxies along the line of sight.  This enables a direct comparison between the simulation and observation. 
The large volume of the lightcone enables more precise quantification of the number of JQ-like systems in the Universe, although this is significantly larger than the 120~arcmin$^2$ covered by the JADES dataset.   
The lightcone is constructed by drawing halos from the SMDPL simulation from the MultiDark suite \cite{Klypin2016} at various redshifts based on their line-of-sight distances. Within these dark matter halos, galaxies are then simulated using the Santa Cruz semi-analytic model (SC SAM) for galaxy formation \cite{Somerville2015,Somerville2021, Yung2019a, Yung2022}, which incorporates the evolution of a wide variety of physical processes, such as cosmological accretion, cooling, star formation, chemical enrichment, and stellar and AGN feedback. 
\editone{The SAM approach does not rely on spatially resolved elements like mass particles and grid cells, but rather utilises analytical and empirical recipes to model the formation and growth of galaxies within halos and their merger histories. Dark matter halos in the lightcone are well-resolved and complete down to $M_{\rm h} \sim 10^{10}\ M_\odot$, which approximately corresponds to a stellar mass of $10^7 \ M_\odot$.}
The performance of the model at high redshift is validated against one-point distribution functions of $M_{\rm UV}$, $M_*$, and SFR \citep{Yung2019a, Yung2019b} and galaxy clustering constraints \citep{Yung2022}. We also note that dust attenuation is accounted for in the predicted photometry of simulated galaxies. 
\editone{However, the lightcone simulation does not employ a prescription for emission lines, which limits us to directly comparing the emission line properties of simulated galaxies and JQ galaxies.}
Sky coordinates of the halos and galaxies therein are computed based on the physical coordinates of sources within the lightcone. 
We refer the readers to refs. \cite{Yung2022,Yung2023} for detailed descriptions for the simulation and the construction of the lightcone. 

\begin{figure}
    \renewcommand{\figurename}{Extended Data Figure}
    \centering
    \includegraphics[width=0.8\linewidth]{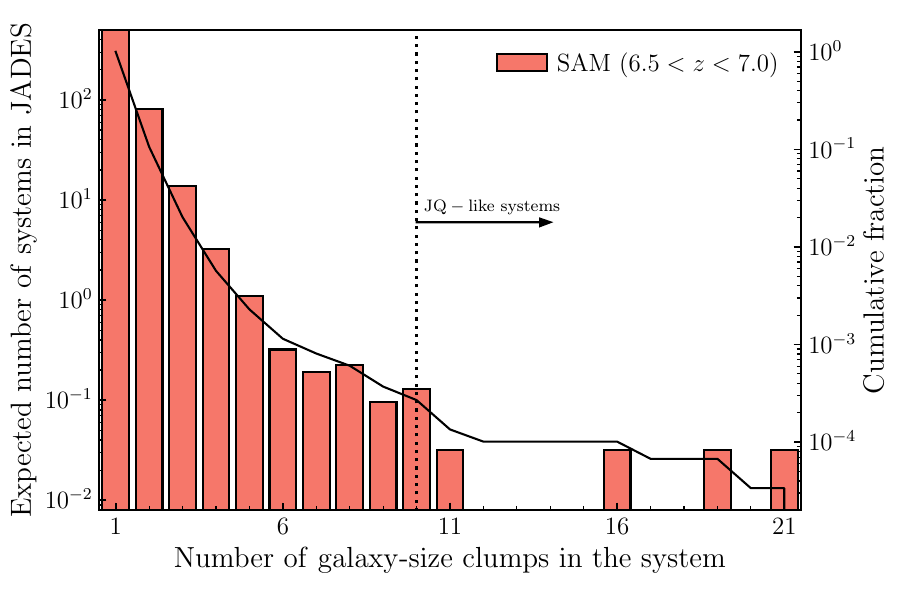}
    \caption{\textbf{Expected distribution of clump numbers in galaxy systems in the JADES survey.} 
    We select the galaxy systems with sizes similar to the JQ from a 2-deg$^2$ cosmological simulation\cite{Yung2023} and rescale them to match the survey volume like JADES. The black solid curve indicates the cumulative fraction of systems as a function of galaxy counts. The black dashed line marks the selection criterion of the JQ-like system in the lightcone.}
    \label{fig:density}
\end{figure}

From the lightcone, we first create a parent galaxy sample at $z=6.5$ -- 7.0 with F277W magnitude $<30$ and then identify a subset of galaxy systems in the parent sample by requiring at least one companion with a velocity difference of $\Delta v<300$ \kms\ and a projected separation of $<5$ arcsec. 
These criteria are chosen to be consistent with the observation of JQ and the depth of the JADES survey.
The redshift range (6.5 -- 7.0) corresponds to a comoving volume of $8.5\times10^6$ cMpc$^3$ for the lightcone simulation.
In total, we select 29,696 unique systems (central and satellite galaxies) from the 2 deg$^2$ lightcone, of which 3154 systems have at least two \editone{galaxies}.
This implies that we would expect $\sim101$ galaxy systems with $>2$ \editone{galaxies} at $z\sim6.7$ -- 7.6 for the JADES survey area.
In Extended Data Figure \ref{fig:density}, we present the expected distribution of clump numbers in galaxy systems in the JADES survey.

\editone{
The SAM tracks the decay of satellite orbits due to dynamical friction and the tidal disruption of infalling satellites (see ref \citep{Somerville2008}, for details), which is capable of reproducing separations that are comparable to the JQ galaxies and clumps.
However, as the SAM approach treats different species of baryonic matter (e.g., hot gas, cold gas, stars) as reservoirs and relies on analytic recipes to track the conversion of mass between different reservoirs, the SAM does not spatially resolve individual star-forming clumps within a star-forming disk or model the clumps that are formed through the merger-induced star formation (for example, merger debris). 
On the other hand, due to the lack of kinematic information of JQ clumps, we cannot distinguish the physical origins of these clumps: some clumps may result from merger-induced star formation, which would not be tracked by the SAM and therefore should be excluded in comparison; others may have formed within companion dark matter halos before merger, in which case they should be included.
Thus, a careful assessment of the number of galaxies in JQ is critical for a fair comparison with the lightcone.}

\editone{
We first consider the galaxy systems and their clumps in JQ that are separated to be the analogs to the galaxies in the lightcone.
Clumps that are closely spaced are considered star-forming regions within the same galaxy.}
These include ELG1 (C1+C2+C3), ELG2 (C5+C6), ELG3 (C8+C9), ELG4 (C10), ELG5 (C11), C4, C7, C12+C13, C14, C16.
We exclude C15 because it is relatively faint compared to the others, and again, we treat ELG3 and ELG4 as separate galaxies due to their distinct stellar populations.
Using this definition (systems with $\geq10$ clumps), we identify 8 such systems in the 2 deg$^2$ lightcone, indicating that JQ has a number density larger than $>99.97\%$ of galaxy systems in the simulation.
\editone{Among these 8 simulated systems, we further select major mergers by requiring a stellar mass ratio between the primary and secondary galaxies to be less than 4 \cite{Duan2025,Puskas2025}.
This yields only 4 major merger systems in the lightcone, corresponding to an expected number of just 0.13 such systems within the JADES survey volume.}

\editone{
To consider a more conservative case, we restrict our comparison to six clearly isolated galaxies/clumps: ELG1 (C1+C2+C3), ELG2 (C5+C6), ELG3 (C8+C9), ELG4 (C10), ELG5 (C11), and C16.
Here, we exclude the clumps surrounding the five main galaxies to test whether the lack of JQ-like systems in the simulation may be due to the omission of clumps formed during merger events.
In this case, we identify 28 systems with $\geq6$ galaxies in the 2 deg$^2$ lightcone, and 11 of which are major merger systems.  
This increases the expected number of JQ-like systems within the JADES survey volume to 0.36, but is still below the observed density.}

\editone{These comparisons indicate that the JQ-like systems appear to be under-represented by the simulated lightcone.
In the SC SAM framework, once a galaxy becomes a satellite, its hot gas supply from cosmological accretion is immediately shut off (under the assumption that the hot gas is added to the reservoir of its respective central galaxy), effectively cutting off the supply of new fuel and causing it to cease star formation prematurely.
Together with the lack of explicitly modeled star-forming clumps, these scenarios may be responsible for this underestimation. 
A recent study comparing the observed galaxy pair fraction with predictions from the SC SAM simulation also found that the simulation underpredicts the pair fraction at $z>6$ \cite{Duan2025}, which is consistent with this interpretation.}

The JQ system was serendipitously discovered in an ELG study based on the JADES F410M image. Given the relatively small survey volume of the JADES F410M image ($\sim120$ arcsec$^2$), we estimate the observed comoving number density of the JQ system to be $3.7^{+3.1}_{-3.0}\times10^{-6}$ cMpc$^{-3}$. The error is estimated based on the Poisson error of the small number statistic \cite{Gehrels1986}.
In comparison, we estimate the comoving number density of JQ-like systems in lightcone simulation to be $9.40\times10^{-7}$ cMpc$^{-3}$.

\subsection{Merging timescale of the JQ system}

Constraining the merging timescale of galaxies is extremely challenging due to significant uncertainties, such as the degeneracy between line-of-sight distance and velocity and the unknown transverse velocity of the galaxies. In systems like JQ, the presence of multiple clumps further complicates precise calculations.
To simplify the estimation, we only consider the merging of two main components in JQ. We define the primary component of JQ as the subsystem of ELG3 and ELG4, the most massive members of JQ.  We define the secondary component as ELG1, the secondary massive galaxy, and the most distant in the projected plane. 
The projected distance between the primary and secondary galaxies is measured based on the projected distance between their most massive clumps, yielding a value of 13.3 kpc.
The total stellar mass is determined by summing the stellar masses of the member clumps. We obtain the stellar mass of $10^{9.32}\ M_\odot$ for the secondary and $10^{9.92}\ M_\odot$ for the primary galaxy. 
To estimate the halo properties of the primary and secondary galaxies, we select the simulated galaxies at $z=6.5$ -- 7.0 from the lightcone catalog \cite{Yung2023} with stellar masses similar to those of our objects.
We identify halos with mass ratios between the primary and secondary galaxies therein to be $\sim 2.3^{+1.2}_{-0.9}$.
The halo properties are determined by the median values of those simulated galaxies, with uncertainties defined by the 16th to 84th percentile from the distribution.
The virial radius and velocity are $29.7^{+5.1}_{-2.9}$ kpc and $227^{+33}_{-20}$ \kms\ for the primary galaxy and $23.0^{+4.2}_{-2.2}$ kpc and $176^{+31}_{-17}$ \kms\ for the secondary galaxy.

We estimate the merging timescales in three methods.  \editone{In the first method, we} estimate the lower limit of the merging timescale, assuming the secondary galaxy follows a circular orbit about the primary galaxy, where dynamical friction dominates its energy and angular momentum losses. 
The merging timescale follows the formula (Equation 12.45 in ref \cite{Mo2010}) based on Chandrasekhar's dynamical friction formula \cite{Chandrasekhar1943}, 
\begin{equation}
    \label{eq:1}
    t_\mathrm{df} = \frac{1.17}{\ln (m_\mathrm{pri}/m_\mathrm{sat})} \left(\frac{r_i}{r_\mathrm{vir}}\right)^2 \left(\frac{m_\mathrm{pri}}{m_\mathrm{sec}}\right) \frac{r_\mathrm{vir}}{V_c},
\end{equation}
where $m_\mathrm{pri}/m_\mathrm{sec}$ is the mass ratio between the primary and secondary galaxies, $V_c$ is the circular velocity, $r_i$ is the distance between the two galaxies, and $r_\mathrm{vir}$ is the virial radius of the primary galaxy.
Adopting the virial velocity of the primary galaxy to be $V_c$ and the projected distance between the two galaxies to be $r_i$, we obtain a merging timescale of $\sim80$ Myr. 
It is important to note that $r_i$ is the projected distance and only represents a lower limit of the true spatial separation. 
If assuming a random orientation of primary and secondary galaxies, the expected deprojected distance would be $\sqrt{3/2}\times13.3 \approx 16.3$ kpc, and the merging timescale increases to $\sim135$ Myr.

\editone{However}, equation \ref{eq:1} can underestimate the merging timescale because, during the merger, the tidal forces of the primary galaxy would gradually strip the matter from the secondary galaxy, decreasing its mass \cite{Somerville1999}.
This effect would increase the merging timescale by a factor of 2 -- 5 \cite{Jiang2008,Mo2010}.
To incorporate the effect of mass loss, we adopt the formula by Ref~\cite{Binney1987} (Equation 8.17) \editone{as the second method}, 
\begin{equation}
    \label{eq:2}
    t = \frac{2.7\ \mathrm{Gyr}}{\ln (m_\mathrm{pri}/m_\mathrm{sat})} \frac{r_i}{30\ \mathrm{kpc}} \left(\frac{V_\mathrm{pri}}{200\ \mathrm{km\ s^{-1}}}\right)^2 \left(\frac{100\ \mathrm{km\ s^{-1}}}{V_\mathrm{sec}}\right)^3,
\end{equation}
where $V_\mathrm{pri}$ and $V_\mathrm{sec}$ are the circular velocity of the two galaxies. 
Adopting the virial velocities of the galaxies as the circular velocities and the projected distance between the two galaxies to be $r_i$, we obtain a merging timescale of $\sim310$ Myr. 
If considering the deprojected distance of $16.3$ kpc, the merging timescale increases to $\sim380$ Myr.

\editone{In the third method}, we adopt a the formula of ref \cite{Jiang2008} (Equation 8), which fits results from an $N$-body numerical simulations, to estimate the merging timescale:
\begin{equation}
    \label{eq:3}
    t = \frac{0.90\epsilon^{0.47} + 0.60}{2C} \left(\frac{m_\mathrm{pri}}{m_\mathrm{sec}}\right) \left(\frac{1}{\ln (1+m_\mathrm{pri}/m_\mathrm{sec})}\right) \left(\frac{\sqrt{r_\mathrm{vir}r_{c}}}{V_{c}}\right),
\end{equation}
where $C$ is a constant, approximately 0.43, and $\epsilon$ is the circularity parameter of the secondary's orbit.
Similar to the previous two models, we assume a circular orbit ($\epsilon=1$) and adopt the virial velocity as $V_c$.
We obtain a merging timescale of $\sim290$ (320) Myr for a distance of 13.3 (16.3) kpc.

Therefore, we conclude the merging timescale of JQ is $\sim$ 100 -- 350 Myr, which spans the range from all the estimates above.

\subsection{No evidence of outflows in ELG1}
\begin{figure}
    \renewcommand{\figurename}{Extended Data Figure}
    \centering
    \includegraphics[width=\linewidth]{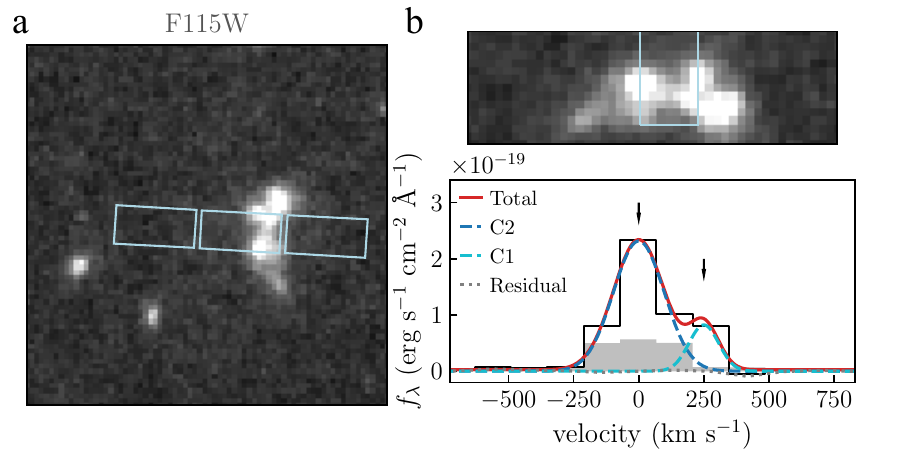}
    \caption{\textbf{NIRSpec spectrum of ELG1.} \textbf{a,} Overlay of NIRSpec MSA shutters (green rectangles) onto the F115W image of ELG1. The shutters simultaneously cover clumps C1 and C2, with the alignment of the two clumps oriented perpendicular to the spatial direction (i.e., parallel to the dispersion direction.) \textbf{b,} The top panel presents the zoom-in to the shutter targeted on ELG1 and the lower panel presents the observed NIRspec spectrum (black). The gray-shaded region indicates the error spectrum. In the lower panel, we also present the best-fitting results using two Gaussian components. The distance between C1 and C2 is $0.18$ arcsec, corresponding to a velocity difference of $250$ \kms. Thus, the centers of the two components are linked based on this velocity difference, as indicated by the two arrows.}
    \label{fig:elg1msa}
\end{figure}

In the bottom right panel of Extended Data Figure \ref{fig:elg1msa}, we present the asymmetric \foiii\ $\lambda5007$ emission line of ELG1 from the NIRSpec Medium Grating observation. 
This asymmetric profile does not indicate the presence of galactic outflows, instead, it is due to the two clumps (C1 and C2) being simultaneously targeted within the same NIRSpec shutter, as shown in Extended Data Figure \ref{fig:elg1msa}a.
The alignment between C1 and C2 is nearly parallel to the dispersion direction, with a separation of $0.18$ arcsec.
This implies a velocity offset of $\sim250$ \kms\ between C1 and C2 in the dispersion direction of the NIRSpec spectrum.
To model their emission line profile, we fit their \foiii\ $\lambda5007$ line with two Gaussian components. The centers of two components are linked based on a velocity difference of $\sim250$ \kms.
We find that the observed \foiii\ $\lambda5007$ profile can be well reproduced by only these two components with widths of $97^{+37}_{-16}$ \kms\ and $79^{+28}_{-21}$ \kms, significantly narrower than the typical outflow width ($>150$ \kms) observed in high-redshift galaxies \cite{Zhang2024} and their local analogs \cite{Peng2025}.
The residual (gray dotted line) is consistent with zero, confirming that the line asymmetry is solely due to the superposition of the two clumps.
Therefore, this suggests that ELG1 lacks strong outflows.

\subsection{Can outflows explain the \foiii+\hb\ bridge between ELG1 and ELG2?}
\begin{figure}
    \renewcommand{\figurename}{Extended Data Figure}
    \centering
    \includegraphics[width=0.5\linewidth]{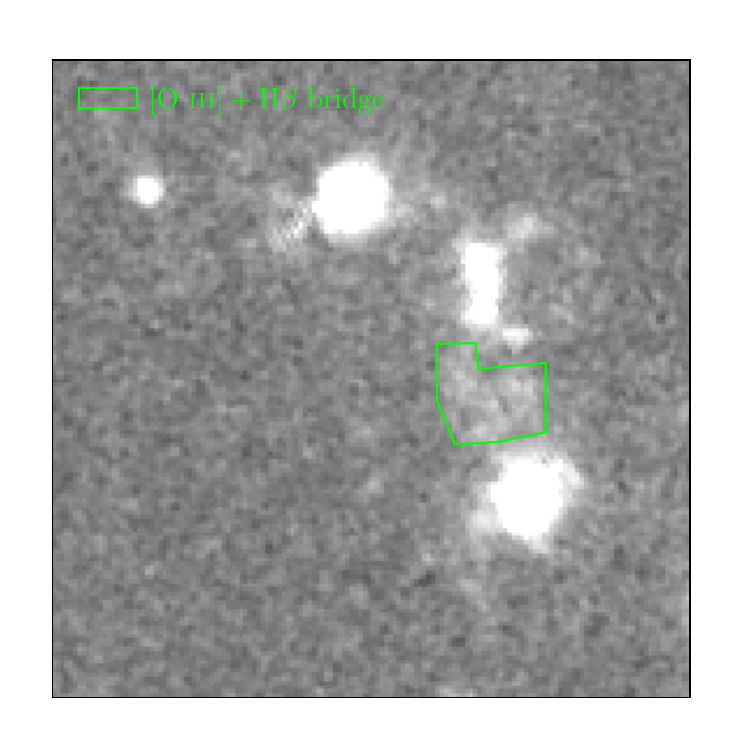}
    \caption{\textbf{Polygonal region (green) used to measure the total \foiii+\hb\ luminosity of \foiii+\hb\ bridge.} The background image is the \foiii+\hb\ map. 
    \editone{This polygonal region is defined to follow the 2-$\sigma$ contour of the \foiii+\hb\ map as shown in Extended Data Figure \ref{fig:oiiimap}, corresponding to a surface brightness of $8.8\times10^{-17}$ erg s$^{-1}$ cm$^{-2}$ arcsec$^{-2}$.  
    The boundaries near ELG1 and ELG2 are adjusted manually to exclude regions associated with the galaxies.}}
    \label{fig:bridge}
\end{figure}

Although the JWST NIRSpec spectrum of ELG1 does not reveal evidence of outflows, this does not conclusively rule out the presence of outflows in other clumps within ELG1 or in ELG2, which were not targeted by the NIRSpec observations.
In this section, we further explore whether the outflows in ELG1 and ELG2 and their adjacent clumps can produce the \foiii+\hb\ bridge between them based on the outflow simulations.
We estimate the \foiii+\hb\ luminosity of \foiii+\hb\ bridge to be the total flux enclosed by a polygonal region defined in Extended Data Figure \ref{fig:bridge} with the contribution of clumps C12 and C13 removed based on Extended Data Table \ref{tab:properties}.
We derive the total observed \foiii+\hb\ luminosity of the bridge to be $\simeq (1.1\pm0.2)\times10^{42}$ erg s$^{-1}$.

We first consider a single-phase galactic wind model, where the direct radiative cooling of hot winds from $\sim10^6 - 10^7$ K to $\sim 10^4 - 10^{5.5}$ K generates the emission lines \cite{Thompson2016,Peng2025}. 
Under this model, we calculate the extended \foiii+\hb\ emission produced by ELG1 and ELG2.
We consider supernova explosions the dominant sources of mass and energy in the hot winds.
Therefore, the energy and mass input rates are proportional to the SFR:
\begin{gather}
    \dot{E}=3\times10^{41}\ \mathrm{erg\ s^{-1}}\ \eta_\mathrm{E}\ \frac{\mathrm{SFR}}{M_\odot\ \mathrm{yr^{-1}}}, \\
    \dot{M} = \eta_\mathrm{M}\ \mathrm{SFR},
\end{gather}
where $\eta_\mathrm{E}$ and $\eta_\mathrm{M}$ are the thermalization-efficiency and mass-loading factors of hot winds, respectively.
Measuring these factors is extremely challenging because the hot winds can only be observed in the X-ray. 
Here, we consider 0.1 and 1 for $\eta_\mathrm{E}$ and consider a wide range of 0.2 -- 1.6 with a step of 0.2 for $\eta_\mathrm{M}$.
These values are consistent with those derived from numerical simulations \cite{Pandya2021,HuC2019,Kim2020}.
The chosen $\eta_\mathrm{E}$ values are also consistent with the X-ray observation of hot winds in the nearby starburst galaxy M82 \cite{Strickland2009}.
For the input SFRs, we adopt those measured from the SED fitting for individual clumps.
We assume a spherical geometry where the galactic winds launch at the half-light radius of galaxies (0.4 -- 0.6 kpc) and expand isotropically. 
The radiative cooling of the hot winds follows the cooling function at $z=6.7$ in ref \cite{Ploeckinger2020}.
The studies of local analogs of high-redshift galaxies revealed that their galactic outflows are usually more metal-enriched \cite{Chisholm2018} than the ISM.  Thus, in this work, we adopt the metallicity of hot winds to be the solar metallicity. 
We utilize the escape velocity to quantify the impact of gravity on the outflow.
The escape velocity relies on the radial distribution of dark matter and is typically larger than the circular velocity by a factor of $>2$ \cite{Binney1987}.
Therefore, we adopt a wide range of 200 -- 500 \kms\ with a step of 100 \kms\ to account for those various uncertainties.
We integrate the luminosities of \foiii\ doublets and \hb\ emission lines from the launch radius to 8 kpc, where the latter corresponds to the distance between ELG1 and ELG2. 
We find the maximal \foiii+\hb\ luminosity (when $\eta_\mathrm{E}=1,\ \eta_\mathrm{M}=1.6$) is only $4.9 \times 10^{39}$ erg s$^{-1}$, 2 -- 3 orders of magnitude lower than the luminosity of \foiii+\hb\ bridge between ELG1 and ELG2.

Recent studies have revealed the multiphase nature of galactic outflows \cite{Veilleux2020, Fisher_2024}. 
The temperature and shear velocity differences between the hot and cold ($\sim 10^4$ K) phases drive the mixing of the two phases and generate the turbulent radiative mixing layers (TRMLs), where the temperature drops into the intermediate regime ($\sim10^5$ K) \cite{Fielding2020,Fielding2022}.
This intermediate temperature regime corresponds to the peak of the radiative cooling function, facilitating the production of \foiii \ and \hb\ emission lines.
Nevertheless, Ref \cite{Peng2025} found that the direct radiative cooling flux of hot winds becomes comparable to ($\gtrsim 10\%$) that of TRMLs when $\eta_{M} \gtrsim 1.0$. 
Additionally, due to the complex interplay between radiative cooling, conduction, and viscous heating in the TRMLs, the cooling flux of TRMLs peaks at $10^4$ K, where the fraction of radiative cooling contributes to \foiii\ becomes relatively small compared to that of hot winds (ratio $\sim 10\%)$. 
As a result of these two factors, the multi-phase galactic wind model with TRMLs \cite{Fielding2022} can only enhance the total \foiii+\hb\ luminosity by $<1$ dex. 
This remains insufficient to explain the luminosity of the \foiii+\hb\ bridge between ELG1 and ELG2. 

Consequently, galactic outflows are unlikely to be the dominant drivers of \foiii+\hb\ bridge.  Therefore we favor the scenario where the \foiii+\hb\ bridge originates from shock-heated, tidally stripped gas.


\backmatter

\bmhead{Correspondence and request for materials} should be addressed to W. Hu.

\bmhead{Acknowledgments}
This research was supported in part by grant NSF PHY-2309135 to the Kavli Institute for Theoretical Physics (KITP). 
This work is based on observations made with the NASA/ESA/CSA {\it JWST}. The data were obtained from the Mikulski Archive for Space Telescopes at the Space Telescope Science Institute, which is operated by the Association of Universities for Research in Astronomy, Inc., under NASA contract NAS 5-03127 for JWST.  This work was supported in part by generous funding provided by Marsha and Ralph Schilling through Texas A\&M University. 

\bmhead{Author contribution} 
W.H. and C.P. designed the layout of this paper. W.H. reduced the data, performed scientific analysis, and wrote the manuscript. C.P. helped with the manuscript writing and scientific analysis.
L.S. helped with the spatially resolved analysis. C.P. and L.Y.A.Y. helped with the lightcone simulation and merging timescale estimation. 
Z.P. performed the outflow analysis based on analytical galactic wind models.
All authors discussed the results and commented on the manuscript.

\bmhead{Conflict of interest/Competing interests}  The authors declare no competing interests.

\bmhead{Data availability} The unprocessed JWST data are available through the Mikulski Archive for Space Telescopes (\url{https://mast.stsci.edu/search/ui/#/jwst}). The reduced JWST images and spectra in this work are publicly available through the STSci High-Level Science Products (\url{https://archive.stsci.edu/hlsp/jades}).
The mock catalogs from the lightcone simulation are publicly available through the Semi-analytic forecasts for the Universe homepage (\url{https://www.simonsfoundation.org/semi-analytic-forecasts}).

\bmhead{Code availability} Codes used in this study are publicly available: Astropy (\url{https://www.astropy.org}), Bagpipes (\url{https://github.com/ACCarnall/bagpipes}), CIGALE (\url{https://cigale.lam.fr}), Pysersic (\url{https://github.com/pysersic/pysersic}), sep (\url{https://github.com/sep-developers/sep}), SExtractor (\url{https://www.astromatic.net/software/sextractor}, and MVT binning (\url{https://github.com/pierrethx/MVT-binning}).

\clearpage


\end{document}